\documentclass[pdftex,twocolumn,epjc3_preprint,runningheads]{svjour3}

\usepackage[colorlinks,citecolor=blue,urlcolor=blue,linkcolor=blue,breaklinks=true]{hyperref}
\usepackage[dvipsnames]{xcolor}

\pdfoutput=1

\usepackage[T1]{fontenc}
\usepackage{lmodern}
\usepackage{calc}
\usepackage{graphicx}
\usepackage{booktabs}
\usepackage{textcomp}
\usepackage{xspace}
\usepackage{relsize}
\usepackage{amssymb}
\usepackage{amsmath}
\usepackage{listings}
\usepackage{microtype}
\usepackage{multirow}
\usepackage{tabularx}
\usepackage{array}
\usepackage{placeins}
\usepackage{cuted}
\usepackage{soul} 
\usepackage{fixltx2e}
\usepackage{slashed}
\usepackage{bm}
\usepackage[numbers,sort&compress]{natbib}
\usepackage[labelfont=bf,font=small]{caption}
\usepackage[skip=-2pt]{subcaption}
\usepackage[clockwise,figuresright]{rotating}
\usepackage{tikz}
\usepackage[normalem]{ulem}
\usepackage[utf8]{inputenc}

\usepackage{etoolbox}
\AfterEndEnvironment{strip}{\leavevmode}

\allowdisplaybreaks

\newcolumntype{L}{>{\raggedright\let\newline\\\arraybackslash\hspace{0pt}}X}
\newcolumntype{R}{>{\raggedleft\let\newline\\\arraybackslash\hspace{0pt}}X}
\newcolumntype{C}{>{\centering\let\newline\\\arraybackslash\hspace{0pt}}X}

\newcommand{\gambitinstitute}[1]{\expandafter\csname #1\endcsname\label{#1}}
\newcommand{\gi}[1]{\gambitinstitute{#1}\and}
\newcommand{\last}[1]{\gambitinstitute{#1}}

\makeatletter

\newcommand{\preprintnumber}[1]{\gdef\@preprintnumber{\begin{flushright}{#1}\end{flushright}}}

\g@addto@macro\bfseries{\boldmath}
\makeatother

\let\underscore\_
\renewcommand{\_}{\discretionary{\underscore}{}{\underscore}}

\makeatletter
\let\orgdescriptionlabel\descriptionlabel
\renewcommand*{\descriptionlabel}[1]{%
  \let\orglabel\label
  \let\label\@gobble
  \phantomsection
  \protected@edef\@currentlabel{#1}%
  \let\label\orglabel
  \orgdescriptionlabel{#1}%
}
\makeatother

\newcommand\cpp[1]{{\lstinline!#1!}}  

\newcommand\yaml[1]{{\lstset{style=yaml}\lstinline!#1!\lstset{style=cpp}}}

\newcommand\term[1]{{\lstset{style=terminal}\lstinline!#1!\lstset{style=cpp}}}
\newcommand\termalt[1]{{\lstset{style=terminalalt}\lstinline!#1!\lstset{style=cpp}}}
\newcommand\fortran[1]{{\lstset{style=fortran}\lstinline!#1!\lstset{style=cpp}}}
\newcommand\py[1]{{\lstset{style=python}\lstinline!#1!\lstset{style=cpp}}}
\newcommand\customtilde{{\raisebox{0.2ex}{\scalebox{0.6}{\boldmath$\sim$}}}}
\newcommand\mathematica[1]{{\lstset{style=Mathematica}\lstinline!#1!\lstset{style=cpp}}}
\newcommand\guminline[1]{{{\lstset{style=gum}\lstinline!#1!}}}
\newcommand\textinline[1]{{{\lstset{style=text}\lstinline!#1!}}}

\def\be{\begin{equation}}
\def\ee{\end{equation}}
\def\ba{\begin{eqnarray}}
\def\ea{\end{eqnarray}}
\newcommand{\bea}{\begin{eqnarray}}
\newcommand{\eea}{\end{eqnarray}}

\lstnewenvironment{lstlistingyaml}{\lstset{style=yaml}}{\lstset{style=cpp}}
\lstnewenvironment{lstlistingterm}{\lstset{style=terminal}}{\lstset{style=cpp}}
\lstnewenvironment{lstlistingfortran}{\lstset{style=fortran}}{\lstset{style=cpp}}
\lstnewenvironment{lstcpp}{\lstset{style=cpp}}{\lstset{style=cpp}}
\lstnewenvironment{lstcppalt}{\lstset{style=cppalt}}{\lstset{style=cpp}}
\lstnewenvironment{lstcppnum}{\lstset{style=cppnum}}{\lstset{style=cpp}}
\lstnewenvironment{lstyaml}{\lstset{style=yaml}}{\lstset{style=cpp}}
\lstnewenvironment{lstgum}{\lstset{style=gum}}{\lstset{style=cpp}}
\lstnewenvironment{lstterm}{\lstset{style=terminal}}{\lstset{style=cpp}}
\lstnewenvironment{lsttermalt}{\lstset{style=terminalalt}}{\lstset{style=cpp}}
\lstnewenvironment{lsttext}{\lstset{style=text}}{\lstset{style=cpp}}
\lstnewenvironment{lstfortran}{\lstset{style=fortran}}{\lstset{style=cpp}}
\lstnewenvironment{lstpy}{\lstset{style=python}}{\lstset{style=cpp}}
\lstnewenvironment{lstmathematica}{\lstset{style=mathematica}}{\lstset{style=cpp}}

\newcommand{\tmpname}{}
\newcommand{\tmplistingname}{}
\makeatletter
\newif\ifATOlabelname
\lst@Key{labelname}{Listing}{\def\ATOlabelname{#1}\global\ATOlabelnametrue}
\makeatother
\lstnewenvironment{lstcpplabel}[1][]{
  \lstset{style=cpp,#1} 
  \ifATOlabelname
    \renewcommand{\tmpname}{\lstlistingname}
    \renewcommand{\tmplistingname}{\lstlistlistingname}
    \renewcommand{\lstlistingname}{\ATOlabelname}
    \renewcommand{\lstlistlistingname}{List of \lstlistingname s}
  \fi
}{
  \renewcommand{\lstlistingname}{\tmpname}
  \renewcommand{\lstlistlistingname}{\tmplistingname}
  \lstset{style=cpp}
}
\definecolor{solarized@base03}{HTML}{002B36}
\definecolor{solarized@base02}{HTML}{073642}
\definecolor{solarized@base01}{HTML}{586e75}
\definecolor{solarized@base00}{HTML}{657b83}
\definecolor{solarized@base0}{HTML}{839496}
\definecolor{solarized@base1}{HTML}{93a1a1}
\definecolor{solarized@base2}{HTML}{EEE8D5}
\definecolor{solarized@base3}{HTML}{FDF6E3}
\definecolor{solarized@yellow}{HTML}{B58900}
\definecolor{solarized@orange}{HTML}{CB4B16}
\definecolor{solarized@red}{HTML}{DC322F}
\definecolor{solarized@magenta}{HTML}{D33682}
\definecolor{solarized@violet}{HTML}{6C71C4}
\definecolor{solarized@blue}{HTML}{268BD2}
\definecolor{solarized@cyan}{HTML}{2AA198}
\definecolor{solarized@green}{HTML}{859900}
\definecolor{darkred}{HTML}{550003}
\definecolor{darkgreen}{HTML}{00AA00}
\definecolor{orchid}{HTML}{AF06F5}

\newcommand\YAMLstringstyle{\footnotesize\color{solarized@green}\mdseries}
\newcommand\YAMLkeystyle{\footnotesize\color{solarized@blue}\ttfamily}
\newcommand\YAMLvaluestyle{\footnotesize\color{blue}\mdseries}
\newcommand\ProcessThreeDashes{\llap{\color{cyan}\mdseries-{-}-}}

\newcommand\CPPcommentstyle{\color{solarized@violet}\footnotesize\ttfamily}
\newcommand\CPPdirectivestyle{\color{solarized@magenta}\footnotesize\ttfamily}
\newcommand\termplainstyle{\footnotesize\ttfamily}

\newcommand\YAMLcommentstyle{\color{solarized@orange}\ttfamily}

\newcommand\processLongMacroDelimiter
{%
\CPPdirectivestyle%
\#define%
}

\lstdefinestyle{cpp}
{
  language=C++,
  basicstyle=\footnotesize\ttfamily,
  basewidth={0.53em,0.44em}, 
  numbers=none,
  tabsize=2,
  breaklines=true,
  escapeinside={@}{@},
  showstringspaces=false,
  numberstyle=\tiny\color{solarized@base01},
  keywordstyle=\color{solarized@orange},
  stringstyle=\color{solarized@red}\ttfamily,
  identifierstyle=\color{solarized@blue},
  commentstyle=\CPPcommentstyle,
  directivestyle=\CPPdirectivestyle,
  emphstyle=\color{solarized@green},
  frame=single,
  rulecolor=\color{solarized@base2},
  rulesepcolor=\color{solarized@base2},
  literate={~} {\customtilde}1,
  moredelim=*[directive]\ \ \#,
  moredelim=*[directive]\ \ \ \ \#
}

\lstdefinestyle{cppalt}
{
  language=C++,
  basicstyle=\footnotesize\ttfamily,
  basewidth={0.53em,0.44em}, 
  numbers=none,
  tabsize=2,
  breaklines=true,
  escapeinside={*@}{@*},
  showstringspaces=false,
  numberstyle=\tiny\color{solarized@base01},
  keywordstyle=\color{solarized@orange},
  stringstyle=\color{solarized@red}\ttfamily,
  identifierstyle=\color{solarized@blue},
  commentstyle=\CPPcommentstyle,
  directivestyle=\CPPdirectivestyle,
  emphstyle=\color{solarized@green},
  frame=single,
  rulecolor=\color{solarized@base2},
  rulesepcolor=\color{solarized@base2},
  literate={~}{\customtilde}1,
  moredelim=**[is][\processLongMacroDelimiter]{BeginLongMacro}{EndLongMacro} 
}

\lstdefinestyle{cppnum}
{
  language=C++,
  basicstyle=\footnotesize\ttfamily,
  basewidth={0.53em,0.44em}, 
  numbers=none,
  tabsize=2,
  breaklines=true,
  escapeinside={@}{@},
  numberstyle=\tiny\color{solarized@base01},
  showstringspaces=false,
  keywordstyle=\color{solarized@orange},
  stringstyle=\color{solarized@red}\ttfamily,
  identifierstyle=\color{solarized@blue},
  commentstyle=\CPPcommentstyle,
  directivestyle=\CPPdirectivestyle,
  emphstyle=\color{solarized@green},
  frame=single,
  rulecolor=\color{solarized@base2},
  rulesepcolor=\color{solarized@base2},
  literate={~} {\customtilde}1,
  moredelim=*[directive]\ \ \#,
  moredelim=*[directive]\ \ \ \ \#
}

\lstdefinestyle{python}
{
  language=Python,
  basicstyle=\footnotesize\ttfamily,
  basewidth={0.53em,0.44em},
  numbers=none,
  tabsize=2,
  breaklines=true,
  escapeinside={@}{@},
  showstringspaces=false,
  numberstyle=\tiny\color{solarized@base01},
  keywordstyle=\color{blue},
  stringstyle=\color{orange}\ttfamily,
  identifierstyle=\color{darkred},
  commentstyle=\color{purple},
  emphstyle=\color{green},
  frame=single,
  rulecolor=\color{solarized@base2},
  rulesepcolor=\color{solarized@base2},
  literate = {~}{\customtilde}1
             {\ as\ }{{\color{blue}\ as\ \color{black}}}3
             {.set}{{\color{black}.}{\color{darkred}set}}4
}

\lstdefinestyle{fortran}
{
  language=Fortran,
  basicstyle=\footnotesize\ttfamily,
  basewidth={0.53em,0.44em},
  numbers=none,
  tabsize=2,
  breaklines=true,
  escapeinside={@}{@},
  showstringspaces=false,
  numberstyle=\tiny\color{solarized@base01},
  keywordstyle=\color{blue},
  stringstyle=\color{orange}\ttfamily,
  identifierstyle=\color{Periwinkle},
  commentstyle=\color{purple},
  emphstyle=\color{green},
  morekeywords={and, or, true, false},
  frame=single,
  rulecolor=\color{solarized@base2},
  rulesepcolor=\color{solarized@base2},
  literate={~}{\customtilde}1
}

\lstdefinestyle{terminal}
{
  language=bash,
  basicstyle=\termplainstyle,
  numbers=none,
  tabsize=2,
  breaklines=true,
  escapeinside={@}{@},
  frame=single,
  showstringspaces=false,
  numberstyle=\tiny\color{solarized@base01},
  keywordstyle=\color{solarized@orange},
  stringstyle=\color{solarized@red}\ttfamily,
  identifierstyle=\color{black},
  commentstyle=\color{solarized@violet},
  emphstyle=\color{solarized@green},
  frame=single,
  rulecolor=\color{solarized@base2},
  rulesepcolor=\color{solarized@base2},
  morekeywords={gambit, cmake, make, mkdir, gum, python, wget, tar, cp, pippi, mpirun},
  deletekeywords={test},
  literate = {/gambit}{{/}{\color{black}}gambit}6
             {gambit/}{{\color{black}}gambit{/}}6
             {gum/}{{\color{black}}gum{/}}4
             {/include}{{/}{\color{black}}include}8
             {cmake/}{{\color{black}}cmake/}6
             {.cmake}{{.}{\color{black}}cmake}6
             {.gum}{{.}{\color{black}}gum}6
             {.tar}{{.}{\color{black}}tar}4
             {source/}{{\color{black}}source{/}}7
             { type}{{\color{black}}{}type}5
             {~}{\customtilde}1
             {math}{{\color{solarized@orange}}math}4
}

\lstdefinestyle{terminalalt}
{
  language=bash,
  basicstyle=\footnotesize\ttfamily,
  numbers=none,
  tabsize=2,
  breaklines=true,
  escapeinside={*@}{@*},
  frame=single,
  showstringspaces=false,
  numberstyle=\tiny\color{solarized@base01},
  keywordstyle=\color{solarized@orange},
  stringstyle=\color{solarized@red}\ttfamily,
  identifierstyle=\color{black},
  commentstyle=\color{solarized@violet},
  emphstyle=\color{solarized@green},
  frame=single,
  rulecolor=\color{solarized@base2},
  rulesepcolor=\color{solarized@base2},
  morekeywords={gambit, cmake, make, mkdir},
  deletekeywords={test},
  literate = {\ gambit}{{\ }{\color{black}}gambit}7
             {/gambit}{{/}{\color{black}}gambit}6
             {gambit/}{{\color{black}}gambit{/}}6
             {/include}{{/}{\color{black}}include}8
             {cmake/}{{\color{black}}cmake/}6
             {.cmake}{{.}{\color{black}}cmake}6
             {~}{\customtilde}1
}

\lstdefinestyle{text}
{
  language={},
  basicstyle=\footnotesize\ttfamily,
  identifierstyle=\color{black},
  numbers=none,
  tabsize=2,
  breaklines=true,
  escapeinside={*@}{@*},
  showstringspaces=false,
  frame=single,
  rulecolor=\color{solarized@base2},
  rulesepcolor=\color{solarized@base2},
  literate={~}{\customtilde}1
}

\lstdefinestyle{yaml}
{
  language=bash,
  escapeinside={@}{@},
  keywords={true,false,null},
  otherkeywords={},
  keywordstyle=\color{solarized@base0}\bfseries,
  basicstyle=\footnotesize\color{black}\ttfamily,
  identifierstyle=\YAMLkeystyle,
  sensitive=false,
  commentstyle=\YAMLcommentstyle,
  morecomment=[l]{\#},
  morecomment=[s]{/*}{*/},
  stringstyle=\YAMLstringstyle\ttfamily,
  moredelim=**[s][\YAMLkeystyle]{,}{:},   
  moredelim=**[l][\YAMLvaluestyle]{:},    
  morestring=[b]',
  morestring=[b]",
  literate =    {---}{{\ProcessThreeDashes}}3
                {>}{{\textcolor{solarized@red}\textgreater}}1
                {gtr}{\textgreater}1
                {grt}{\textgreater}1
                {|}{{\textcolor{solarized@red}\textbar}}1
                {\ -\ }{{\mdseries\color{black}\ -\ \negmedspace}}3
                {\}}{{{\color{black} \}}}}1
                {\{}{{{\color{black} \{}}}1
                {[}{{{\color{black} [}}}1
                {]}{{{\color{black} ]}}}1
                {~}{\customtilde}1,
  breakindent=0pt,
  breakatwhitespace,
  columns=fullflexible
}

\lstdefinestyle{gum}
{
  language=bash,
  escapeinside={@}{@},
  keywords={true,false,null,all},
  otherkeywords={},
  keywordstyle=\color{solarized@base02}\bfseries,
  basicstyle=\footnotesize\color{black}\ttfamily,
  identifierstyle=\color{solarized@magenta},
  sensitive=false,
  commentstyle=\color{solarized@cyan}\ttfamily,
  morecomment=[l]{\#},
  morecomment=[s]{/*}{*/},
  stringstyle=\footnotesize\color{solarized@base01}\mdseries\ttfamily,
  moredelim=**[l][\footnotesize\color{solarized@base02}\mdseries]{:},    
  morestring=[b]',
  morestring=[b]",
  literate =    {---}{{\ProcessThreeDashes}}3
                {grt}{{\textcolor{solarized@magenta}\textgreater}}1
                {gtr}{{\textcolor{solarized@base02}\textgreater}}1
                {/>}{{\textcolor{solarized@magenta}\textgreater}}1
                {/<}{{\textcolor{solarized@magenta}\textless}}1
                {lss}{{\textcolor{solarized@base02}\textless}}1
                {pls}{{\textcolor{solarized@magenta}+}}1
                {mns}{{\textcolor{solarized@magenta}-}}1
                {|}{{\textcolor{solarized@base02}\textbar}}1
                {\ -\ }{{\mdseries\color{solarized@base02}\ -\ \negmedspace}}3
                {\}}{{{\color{solarized@base02} \}}}}1
                {\{}{{{\color{solarized@base02} \{}}}1
                {[}{{{\color{solarized@base02} [}}}1
                {]}{{{\color{solarized@base02} ]}}}1
                {~}{\customtilde}1,
  breakindent=0pt,
  breakatwhitespace,
  columns=fullflexible
}

\lstdefinestyle{mathematica}
{
  language={Mathematica},
  basicstyle=\footnotesize\ttfamily,
  basewidth={0.53em,0.44em},
  numbers=none,
  tabsize=2,
  breaklines=true,
  postbreak=,
  escapeinside={@}{@},
  numberstyle=\tiny\color{black},
  showstringspaces=false,
  numberstyle=\tiny\color{solarized@base01},
  keywordstyle=\color{solarized@orange},
  stringstyle=\color{solarized@red}\ttfamily,
  identifierstyle=\color{solarized@orange}\ttfamily,
  commentstyle=\color{solarized@gray}\ttfamily,
  directivestyle=\color{solarized@orange}\ttfamily,
  emphstyle=\color{solarized@green},
  frame=single,
  rulecolor=\color{solarized@base2},
  rulesepcolor=\color{solarized@base2},
  literate={~} {\customtilde}1,
  moredelim=*[directive]\ \ \#,
  moredelim=*[directive]\ \ \ \ \#,
  mathescape=false
}

\newcommand{\doublecross}[2]{\hyperref[#2]{\textbf{#1}}}
\newcommand{\doublecrosssf}[2]{\hyperref[#2]{\textbf{\textsf{#1}}}}

\newcommand{\startglossary}{\section{Glossary}\label{glossary}Here we explain some terms that have specific technical definitions in \GB.\begin{description}}
\newcommand{\finishglossary}{\end{description}}

\newcommand{\bcode}{\begin{lstlisting}}
\newcommand{\ecode}{\end{lstlisting}}

\newcommand{\gambit}{\textsf{GAMBIT}\xspace}

\newcommand{\darkbit}{\textsf{DarkBit}\xspace}
\newcommand{\colliderbit}{\textsf{ColliderBit}\xspace}

\newcommand{\GB}{\gambit}

\newcommand{\pythia}{\textsf{Pythia}\xspace}

\newcommand{\mo}{\micromegas}
\newcommand{\micromegas}{\textsf{micrOMEGAs}\xspace}

\newcommand\gamLike{\textsf{gamLike}\xspace}
\newcommand\gamlike{\gamLike}

\newcommand{\gum}{\textsf{GUM}\xspace}

\newcommand{\CH}{\textsf{CalcHEP}\xspace}

\newcommand\beq{\begin{equation}}
\newcommand\eeq{\end{equation}}

\newcommand{\subparagraph}{} 
\journalname{Eur.\ Phys.\ J.\ C}
\smartqed

\usepackage[title]{appendix}

\usepackage{amssymb}
\usepackage{pifont}

\usepackage{natbib}
\usepackage{graphicx}
\usepackage{float}
\usepackage{scalerel}

\usepackage{multirow}
\usepackage{booktabs}
\usepackage{wrapfig}

\usepackage{tikz-feynman}
\tikzfeynmanset{compat=1.1.0}

\newcommand{\madanalysis}{\textsf{MadAnalysis}\xspace}

\begin{document}

\preprintnumber{gambit-physics-2023, TTP23-007, P3H-23-012, ADP-23-07/T1216}

\title{Global fits of simplified models for dark matter with GAMBIT}
\subtitle{II. Vector dark matter with an $s$-channel vector mediator}

\author{\mbox{Christopher Chang\thanksref{uq,a} \and
 Pat Scott\thanksref{qb} \and
 Tom\'as~E.~Gonzalo\thanksref{kitTTP} \and
 Felix Kahlhoefer\thanksref{kitTTP} \and
 Martin White\thanksref{adelaide}}
}

\institute{
  \gi{uq}
  \gi{qb}
  \gi{kitTTP}
  \last{adelaide}
}

\thankstext{a}{christopher.chang@uq.net.au}

\titlerunning{Global fits of simplified DM models II. Vector DM with $s$-channel vector mediator}
\authorrunning{Chang et al.}

\date{Received: date / Accepted: date}


\maketitle

\begin{abstract}

Global fits explore different parameter regions of a given model and apply constraints obtained at many energy scales. This makes it challenging to perform global fits of simplified models, which may not be valid at high energies. In this study, we derive a unitarity bound for a simplified vector dark matter model with an $s$-channel vector mediator and apply it to global fits of this model with \GB in order to correctly interpret missing energy searches at the LHC. Two parameter space regions emerge as consistent with all experimental constraints, corresponding to different annihilation modes of the dark matter. We show that although these models are subject to strong validity constraints, they are currently most strongly constrained by measurements less sensitive to the high-energy behaviour of the theory. Understanding when these models cannot be consistently studied will become increasingly relevant as they are applied to LHC Run 3 data.

\end{abstract}

\tableofcontents


\section{Introduction}

As successful a theory as the Standard Model (SM) has been, there are many reasons for expecting it to exist within an even more descriptive particle theory. One of these reasons for beyond-Standard Model (BSM) physics is a number of astrophysical and cosmological observations that may require additional unseen matter~\cite{Zwicky33, bullet, wmap3year}. The WIMP hypothesis postulates that this matter consists of a Weakly-Interacting Massive Particle, and is a popular theory as it may explain the observed cosmological relic abundance of dark matter (DM)~\cite{Lee:1977ua} and be strongly constrained by near-future experiments~\cite{Arcadi:2017kky}.

WIMP candidates are present in many UV-complete theories including supersymmetric and extra-dimensional models. Rather than focus on these UV-complete theories, this study will instead focus on a simplified model. These are a class of effective theories where the particle that mediates interactions between DM and SM particles is explicitly included. In the limit of large mediator masses, the traditional DM effective theory is recovered. These models have been reviewed in detail in many works, including Refs.~\cite{Abdallah:2015ter,Arina:2018zcq,DeSimone:2016fbz,Albert:2017onk,Boveia:2016mrp,Kahlhoefer:2017dnp,Arcadi:2017kky, Morgante:2018tiq}. They have become the preferred method for modelling the simultaneous impact of low and high energy probes~\cite{DEramo:2016gos, Carpenter:2016thc, Abercrombie:2015wmb}. Studies of these models are often grouped to include multiple simplified models with different mediator and DM spins. This work will instead focus on a single model, in which a vector DM candidate interacts with a vector mediator in the $s$-channel. Details of this model are discussed in section~\ref{Models}. For global fits of models with scalar or fermion DM candidates, we refer the reader to the previous work in this series~\cite{DMSimpI}.

Models containing new vector particles can come with additional theoretical challenges in the high energy limit of the theory, arising from the requirement of unitarity of the scattering matrix. Unitarity violation is a sign that the theory must be extended for it to be theoretically consistent; for example, unitarity violation in SM gauge boson interactions gave one of the early theoretical limits on the mass of the Higgs boson~\cite{Lee:1977yc}. Likewise, unitarity arguments have been used to place an upper bound on the mass of DM particles that obtain their relic abundance through thermal freeze-out~\cite{Griest:1989wd}.

Vector DM simplified models have been studied in detail for both high and low energy experiments. For direct detection constraints, it has been shown that additional non-relativistic effective operators may arise in these models \cite{Dent:2015zpa,Catena:2019hzw}, and that the use of polarized targets may distinguish between fermion and vector DM candidates \cite{Catena:2018uae}. Assuming a detection of signal events at the XENONnT experiment, prospects for finding these models during Run III of the Large Hadron Collider (LHC) in dijet searches \cite{Baum:2018lua} and mono-jet searches \cite{Baum:2017kfa} have been studied along with relic density limits \cite{Catena:2017xqq}.

In this work, we derive a unitarity bound from the self-scattering of vector DM and show the similarity in constraint between this and the requirement of a physical decay width of the mediator. We follow this with a global fit of this model using \GB \textsf{v2.4}, including the decay width and unitarity requirements in our calculations.

This paper is structured as follows. In Section \ref{Models}, we describe the simplified model that we study, and the reasons behind the choice of couplings. In section \ref{Unitarity Violation}, we derive a unitarity bound on this model. Section \ref{Constraints} describes the set of experimental constraints we use to perform a global fit of this model and section \ref{Results} provides our results. Finally, Section \ref{section:discussion} briefly discusses the potential to observe these particles at near-future experiments and presents our conclusions. The samples from our scans, the corresponding \GB plotting scripts and a detailed unitarity bound proof can be downloaded from Zenodo~\cite{Zenodo_DMsimpII}.

\section{Model}
\label{Models}

The general form of the Lagrangian for a simplified model of vector DM $X^\mu$ coupled to quarks via a mediator $V^\mu$ with vector and axial-vector couplings is \cite{Baum:2017kfa}
\begin{align}
\label{ModelEqFull}
\begin{split}
\mathcal{L}_{\text{BSM}} = & -\frac{1}{2} X_{\mu \nu}^{\dagger} X^{\mu \nu} + m_{\text{DM}}^2 X^{\dagger}_{\mu} X^{\mu} - \frac{1}{4} F^{\prime}_{\mu \nu} F^{\prime \mu \nu} \\ & - \frac{1}{2} m_{\text{M}}^2 V_{\mu} V^{\mu} -  h_{3} V_{\mu} \bar{q} \gamma^{\mu} q -   h_{4} V_{\mu} \bar{q} \gamma^{\mu} \gamma^{5} q \\
 & - \frac{\lambda_{\text{DM}}}{2}(X^{\dagger}_{\mu} X^{\mu})^2 - \frac{\lambda_{\text{M}}}{4} (V_{\mu}V^{\mu})^2 \\
 & - \frac{b_3}{2} V_{\mu}^2 (X^{\dagger}_{\nu} X^{\nu}) -  \frac{b_4}{2} (V^{\mu} V^{\nu}) (X^{\dagger}_{\mu} X_{\nu}) \\
 & - [i b_5 X^{\dagger}_{\nu} \partial_{\mu} X^{\nu} V^{\mu} +  b_6 X^{\dagger}_{\mu} \partial^{\mu} X_{\nu} V^{\nu} \\
  & \qquad +  b_7 \epsilon_{\mu \nu \rho \sigma} (X^{\dagger \mu} \partial^{\nu} X^{\rho}) V^{\sigma} + \text{h.c.}]\,,
\end{split}
\end{align}
where \(X_{\mu \nu}\) is the field strength tensor for the vector DM, and \(F^{\prime}_{\mu \nu}\) for the mediator.

To reduce the complexity of this simplified model and the dimensionality of the corresponding parameter space, we make a number of simplifying assumptions. First, we neglect any four-field interactions, which are expected to be irrelevant for phenomenology, and therefore set the couplings $\lambda_{\text{DM}}$, $\lambda_{\text{M}}$, $b_3$ and $b_4$ to zero. Furthermore, we assume that the simplified model conserves $CP$ symmetry, which requires the real components of $b_6$ and $b_7$ in eq.~\eqref{ModelEqFull} to vanish. Finally, to preserve the SM gauge structure, we concentrate on vector-like couplings of the mediator to SM quarks and set $h_4 = 0$.

With these restrictions, one finds that the imaginary components of $b_6$ and $b_7$ only give rise to interactions that vanish in the limit of zero momentum transfer, leading to strongly suppressed constraints from direct detection experiments. Including these couplings in our global fits would therefore lead to rather trivial results, while at the same time requiring significant additional work in order to correctly treat the non-relativistic effective operators $\mathcal{O}_{19}$ and $\mathcal{O}_{20}$ introduced in Ref.~\cite{Catena:2019hzw} and the interference between different operators in the simulation of LHC events. We therefore neglect these couplings in the present work and focus on the two interaction terms proportional to $h_3$ and $b_5$.

Therefore, the Lagrangian of the model we adopt is
\begin{align}
\begin{split}
\mathcal{L}_{\text{BSM}} = & -\frac{1}{2} X_{\mu \nu}^{\dagger} X^{\mu \nu} + m_{\text{DM}}^2 X^{\dagger}_{\mu} X^{\mu} \\
- & \frac{1}{4} F^{\prime}_{\mu \nu} F^{\prime \mu \nu} - \frac{1}{2} m_{\text{M}}^2 V_{\mu} V^{\mu} +   g_{\text{q}} V_{\mu} \bar{q} \gamma^{\mu} q \\
 - & i g_{\text{DM}} \Big(X^{\dagger}_{\nu} \partial_{\mu} X^{\nu} - (\partial_{\mu} X^{\dagger \nu}) X_{\nu} \Big) V^{\mu}\,,
\end{split}
\end{align}
where we choose to label the quark coupling as $g_{\text{q}}$ and the DM coupling as $g_{\text{DM}}$ to agree with our previous work~\cite{DMSimpI}. Both couplings can be taken as purely real since any imaginary phase can be absorbed into a redefinition of the fields.

Perturbative unitarity breaks down in large regions of the parameter space of this model due to the poor high energy behaviour of the longitudinal polarized modes of the vector DM. Following the same approach as Ref.~\cite{Kahlhoefer:2015bea}, here we derive an approximate unitarity bound for this model in terms of the Mandelstam variable $s$, from scattering of vector DM
\begin{align}
    s \lesssim \frac{\sqrt{48 \pi} m_{\text{DM}}^2}{g_{\text{DM}}}\,.
\end{align}
Section \ref{Unitarity Violation} derives this relation, and section \ref{constraints:monojet} describes how unitarity was imposed on simulated collider events in our global scan. In Appendix~\ref{section:appendix}, we present the equivalent bound if the $b_{6}$ and $b_{7}$ couplings of eq.~\eqref{ModelEqFull} are included alongside the $b_{5}$ coupling.

The onshell decay width of the mediator to a pair of DM particles, $V \to X X$, is
\begin{align}
\begin{split}
    \mathrm{\Gamma}_{\text{DM}} =  & \frac{g_{\text{DM}}^2 \sqrt{1 - \frac{4 m_{\text{DM}}^2}{m_{\text{M}}^2} }}{192 \pi m_{\text{M}} m_{\text{DM}}^4} \times \\
    &  \Big( m_{\text{M}}^6 - 8 m_{\text{M}}^4 m_{\text{DM}}^2 + 28 m_{\text{M}}^2 m_{\text{DM}}^4 - 48 m_{\text{DM}}^6 \Big)\,,
\end{split}
\end{align}
and the width to a given pair $i$ of SM quarks, $V \to q_i q_i$, is
\begin{align}
    \mathrm{\Gamma}_{\text{q}_i} = \frac{g_{\text{q}}^2 \sqrt{1 - \frac{4 m_{\text{q}_i}^2}{m_{\text{M}}^2} }}{4 \pi m_{\text{M}}}  \Big( m_{\text{M}}^2 + 2 m_{\text{q}_i}^2 \Big)\,.
\end{align}
The total width of the mediator should not exceed the mediator mass, or else the perturbative description of DM interactions via mediator exchange is expected to break down.

\section{Unitarity Violation}
\label{Unitarity Violation}

\subsection{Forming Unitarity constraints from partial waves}

Unitarity bounds are formed from partial wave analysis of the scattering of vector DM particles. For examples on the use of this method, see e.g.\ Refs.\ \cite{Kahlhoefer:2015bea,Lee:1977yc,Chanowitz:1978mv}. From the requirement of partial wave unitarity, the scattering amplitude must obey the bounds
\begin{align}
    0 \leq \text{Im}(\mathcal{M}^{J}_{ii}) \leq 1\,,
\end{align}
and
\begin{align}
\label{UnitarityRequirement}
    |\text{Re}(\mathcal{M}^{J}_{ii})| \leq \frac{1}{2}\,.
\end{align}
Here $\mathcal{M}^{J}_{ii}$ is the full scattering matrix element between 2-particle states $i$ where the initial and final state particles are the same (hence the repeated index $i$), for the $J$th partial wave. Tree-level amplitudes are generally used to form these bounds, assuming that the higher orders do not provide significant corrections to the amplitude. In this way, the resulting bound may be interpreted as a ``perturbative unitarity'' bound. In the case of zero initial and final total spin,
\begin{align}
\label{UnitarityIntegral}
    \mathcal{M}^{J}_{ii}(s) = \frac{1}{32 \pi} \beta_{ii} \int_{-1}^{1} d \cos{\theta} P^J(\cos{\theta}) \mathcal{M}_{ii}(s,\cos{\theta})\,.
\end{align}
Here \(P^J(x)\) is the Legendre polynomial of order $J$, $\theta$ is the scattering angle and $s$ is the square of the centre-of-mass energy. An additional factor of $1/\sqrt{2}$ must be applied to the right hand side for each initial or final state with identical particles. The term $\beta_{ii}$ is a kinematic factor, which for a final state of equal mass DM particles becomes
\begin{align}
\begin{split}
  \beta_{ii} = &\frac{\sqrt{(s-(m_1 + m_2)^2)(s- (m_1 - m_2)^2)}}{s} \\
   = &\sqrt{\frac{s - 4 m_{\text{DM}}^2}{s}}\,.
\end{split}
\end{align}
In the high-energy limit ($s \rightarrow \infty$), $\beta_{ii}$ approaches 1. As the zeroth order usually dominates, it is often sufficient to study
\begin{align}
\label{Eq:UnitarityIntegral}
    \mathcal{M}^{0}_{ii}(s) = \frac{\beta_{ii}}{64 \pi} \int_{-1}^{1} d \cos{\theta} \mathcal{M}_{ii}(s,\cos{\theta})\,.
\end{align}
In the following derivation, we consider the self-scattering of DM, rather than DM with its antiparticle.  The particle-antiparticle scattering via $s$-channel mediator exchange will also face poor behaviour at high energies, however this will be effectively covered anyway by our additional requirement that the perturbative description of the off-shell decay width of the mediator (including to DM particle-antiparticle pairs) does not break down.


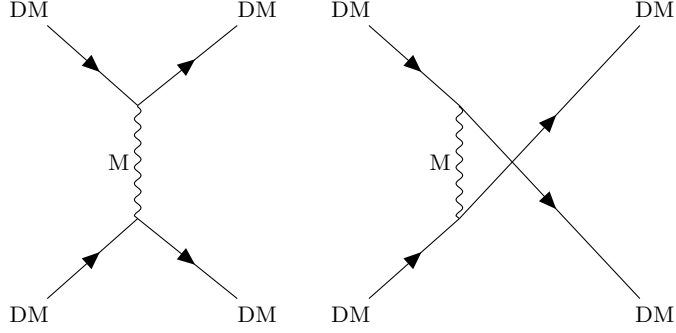
\begin{figure*}[tbp]
  \centering
  \begin{tikzpicture}
  \begin{feynman}
    \vertex (a);
    \vertex [below      =of a] (b);
    \vertex [above left=of a] (i1) {DM};
    \vertex [below left =of b] (i2) {DM};
    \vertex [right      =3cm of i2] (f1) {DM};
    \vertex [right      =3cm of i1] (f2) {DM};

    \diagram* {
      (i1) -- [fermion] (a)
           -- [fermion] (f2),
      (a) -- [photon, edge label'=M] (b),
      (i2) -- [fermion] (b),
      (b) -- [fermion] (f1),
    };
  \end{feynman}
  \end{tikzpicture}
  \hspace*{1em}
  \begin{tikzpicture}
  \begin{feynman}
    \vertex (a);
    \vertex [below      =of a] (b);
    \vertex [above left=of a] (i1) {DM};
    \vertex [below left =of b] (i2) {DM};
    \vertex [right      =4cm of i2] (f1) {DM};
    \vertex [right      =4cm of i1] (f2) {DM};

    \diagram* {
      (i1) -- [fermion] (a)
           -- [fermion] (f1),
      (a) -- [photon, edge label'=M] (b),
      (i2) -- [fermion] (b),
      (b) -- [fermion] (f2),
    };
  \end{feynman}
  \end{tikzpicture}
  \caption{$t$-channel (left) and $u$-channel (right) diagrams relevant for perturbative unitarity bounds. }
  \label{TandUFeynMannDiagrams}
\end{figure*}

The tree level amplitude of DM self-scattering has contributions from $t$ and $u$ channel processes (see Figure~\ref{TandUFeynMannDiagrams}), which can be derived separately, and summed together. This is most easily understood in the centre of mass frame, where for incoming particles (with momenta $p_{(1)}$ and $p_{(2)}$) and outgoing particles (with momenta $p_{(3)}$ and $p_{(4)}$),
\begin{align}
\begin{split}
    & p_{(1)} = \big(\text{E},0,0,\text{P}\big) \\
    & p_{(2)} = \big(\text{E},0,0,-\text{P}\big) \\
    & p_{(3)} = \big(\text{E}, \text{P} \sin{\theta} ,0,\text{P} \cos{\theta} \big) \\
    & p_{(4)} = \big(\text{E},-\text{P} \sin{\theta} ,0,- \text{P}\cos{\theta} \big)\,.
\end{split}
\end{align}
Here \(\text{E} = \frac{\text{E}_{\text{cm}}}{2}\) is the incoming particle energy and P is the magnitude of the incoming momentum of each particle. The longitudinal polarisations will most strongly violate unitarity, and so it is sufficient to solely form a bound from evaluating the amplitude for incoming longitudinally polarised DM particles. In the centre of mass frame, these are
\begin{align}
\begin{split}
    & \epsilon_{(1)} = \frac{1}{m_{\text{DM}}}\big(\text{P},0,0,\text{E}\big) \\
    & \epsilon_{(2)} = \frac{1}{m_{\text{DM}}}\big(\text{P},0,0,-\text{E}\big) \\
    & \epsilon_{(3)} = \frac{1}{m_{\text{DM}}}\big(\text{P},\text{E} \sin{\theta},0,\text{E} \cos{\theta}\big) \\
    & \epsilon_{(4)} = \frac{1}{m_{\text{DM}}}\big(\text{P},-\text{E} \sin{\theta},0,-\text{E} \cos{\theta}\big)\,.
\end{split}
\end{align}
The amplitude for $t$-channel DM-DM scattering at tree-level is
\begin{align}
\begin{split}
    \mathcal{M} = &\frac{g_{\text{DM}}^2}{k_{(t)}^2 - m_{\text{M}}^2}\Big[\epsilon_{(3) \tau} (p_{(1) \mu} + p_{(3) \mu}) \epsilon^{\phantom{i}\tau}_{(1)} \Big] \\
    & \times \Big( g^{\mu \nu} - \frac{k_{(t)}^{\mu} k_{(t)}^{\nu}}{m_{\text{M}}^2}\Big) \Big[ \epsilon_{(2) \sigma} (p_{(2) \nu} + p_{(4) \nu}) \epsilon_{(4)}^{\phantom{i}\sigma}\Big]\,,
\end{split}
\end{align}
where
\begin{align}
\begin{split}
    k_{(t)} =&  p_{(1)} - p_{(3)} = p_{(4)} - p_{(2)} \\
    =& \big(0,-\mathrm{P} \sin{\theta}, 0, \mathrm{P}(1 - \cos{\theta})\big)\,.
\end{split}
\end{align}
Evaluating this amplitude in the centre of mass frame gives
\begin{align}
\begin{split}
    \mathcal{M}_t = & \frac{-g_{\text{DM}}^2}{m_{\text{DM}}^4 \big(2\mathrm{P}^2(1-\cos{\theta}) + m_{\text{M}}^2\big)} \\
    & \times\Big(\mathrm{P}^2 - (\mathrm{P}^2 + m_{\text{DM}}^2) \cos{\theta} \Big)^2 \\
    & \times  \Big(6\mathrm{P}^2 + 4 m_{\text{DM}}^2 + 2\mathrm{P}^2 \cos{\theta} \Big)\,.
\end{split}
\end{align}
Similarly, the scattering amplitude for $u$-channel DM DM scattering at tree-level is
\begin{align}
\begin{split}
    \mathcal{M}_u = & \frac{-g_{\text{DM}}^2}{m_{\text{DM}}^4 \big(2\mathrm{P}^2(1+\cos{\theta}) + m_{\text{M}}^2\big)} \\
    & \times\Big(\mathrm{P}^2 + (\mathrm{P}^2 + m_{\text{DM}}^2) \cos{\theta} \Big)^2 \\
    & \times  \Big(6\mathrm{P}^2 + 4 m_{\text{DM}}^2 - 2\mathrm{P}^2 \cos{\theta} \Big)\,.
\end{split}
\end{align}

\subsection{Unitarity Bound}
\label{sec:UnitarityBound}

The total amplitude of the scattering process is
%
%
%
%
%
%
\begin{align}
    \mathcal{M}_{ii}(s,\cos{\theta}) = \mathcal{M}_t + \mathcal{M}_u\,.
\end{align}
Performing the integral in eq.~\eqref{Eq:UnitarityIntegral} and substituting into eq.~\eqref{UnitarityRequirement} gives the bound on parameters to satisfy unitarity
\begin{align}
\label{UnitatityBoundEq}
\begin{split}
    \Bigg|\frac{g_{\text{DM}}^2}{96 \pi m_{\text{DM}}^4 (\frac{s}{2} - 2 m_{\text{DM}}^2)^3} \sqrt{\frac{s - 4 m_{\text{DM}}^2}{s}} \\
    \Bigg( 2 \Big((\frac{s}{2} - 2 m_{\text{DM}}^2)^2 \big(\frac{s^2}{4} ( m_{\text{DM}}^2 - s )\\
    + \frac{3 m_{\text{DM}}^2 s}{4}(3s - 4 m_{\text{DM}}^2) + 3m_{\text{DM}}^4 (\frac{s}{2} - 2m_{\text{DM}}^2) \big) \\
    + \frac{3 m_{\text{M}}^2 s}{8} (s -4  m_{\text{DM}}^2)  (\frac{3s^2}{8} + \frac{1}{2} m_{\text{DM}}^2 s - 4m_{\text{DM}}^4) \\
    + \frac{3m_{\text{M}}^4 s^2}{32} (s - 4 m_{\text{DM}}^2) \Big)\\
    - 3 \big(\frac{m_{\text{M}}^2 s}{4}+m_{\text{DM}}^2 (\frac{s}{2} - 2 m_{\text{DM}}^2) \big)^2 \\
     \big(2s - 4 m_{\text{DM}}^2+m_{\text{M}}^2  \big) \\
     \ln{\Big(\frac{s - 4 m_{\text{DM}}^2+m_{\text{M}}^2}{m_{\text{M}}^2} \Big)}  \Bigg) \Bigg| \leq \frac{1}{2}\,.
\end{split}
\end{align}
Since unitarity is increasingly violated as the collision energy increases, the limit $s \gg m_{\text{DM}}^2$ is often taken in the literature. If this limit is taken, this bound simplifies to
\begin{align}
    s \lesssim \frac{\sqrt{48 \pi} m_{\text{DM}}^2}{g_{\text{DM}}}\,.
\end{align}
The validity of this limit breaks down for small DM masses and large couplings. In these cases, the complete bound eq.~\eqref{UnitatityBoundEq} should be used.

Even though the unitarity requirement above has been derived for the case of DM self-scattering, the resulting bound can be interpreted more generally as the energy scale where the interactions between DM particles and the vector mediator become unphysical. We will therefore apply the unitarity bound from eq.~\eqref{UnitatityBoundEq} to any process in which a pair of DM particles is produced, with $\sqrt{s}$ being replaced by the invariant mass of the DM pair $m_\text{inv}$. In particular, this requirement will be implemented in our simulation of LHC monojet events (see section~\ref{Constraints}), where we will discard any event that violates the unitarity bound. In other words, we apply LHC constraints only on those regions of phase space where the simplified model predictions can be trusted, and set conservative bounds otherwise.

It is worth noting that for $m_\text{DM} < m_\text{M}/2$, we expect mono-jet production to proceed dominantly via an on-shell mediator, such that $m_\text{inv} \approx m_\text{M}$. Hence, for
\begin{align}
    m_\text{M}^2 \gtrsim \frac{\sqrt{48 \pi} m_{\text{DM}}^2}{g_{\text{DM}}} \; ,
\end{align}
virtually all events will be removed by the unitarity requirement such that the LHC mono-jet bounds are effectively absent. However, parameter points in this region typically also violate the requirement on the decay width from eq.~\eqref{eq:width}, such that they would be excluded from the analysis anyway.

\subsection{Physical Decay Widths}
\label{sec:DecayUnitarityComparison}

\begin{figure}[tbp]
  \centering
  \includegraphics[width=\columnwidth]{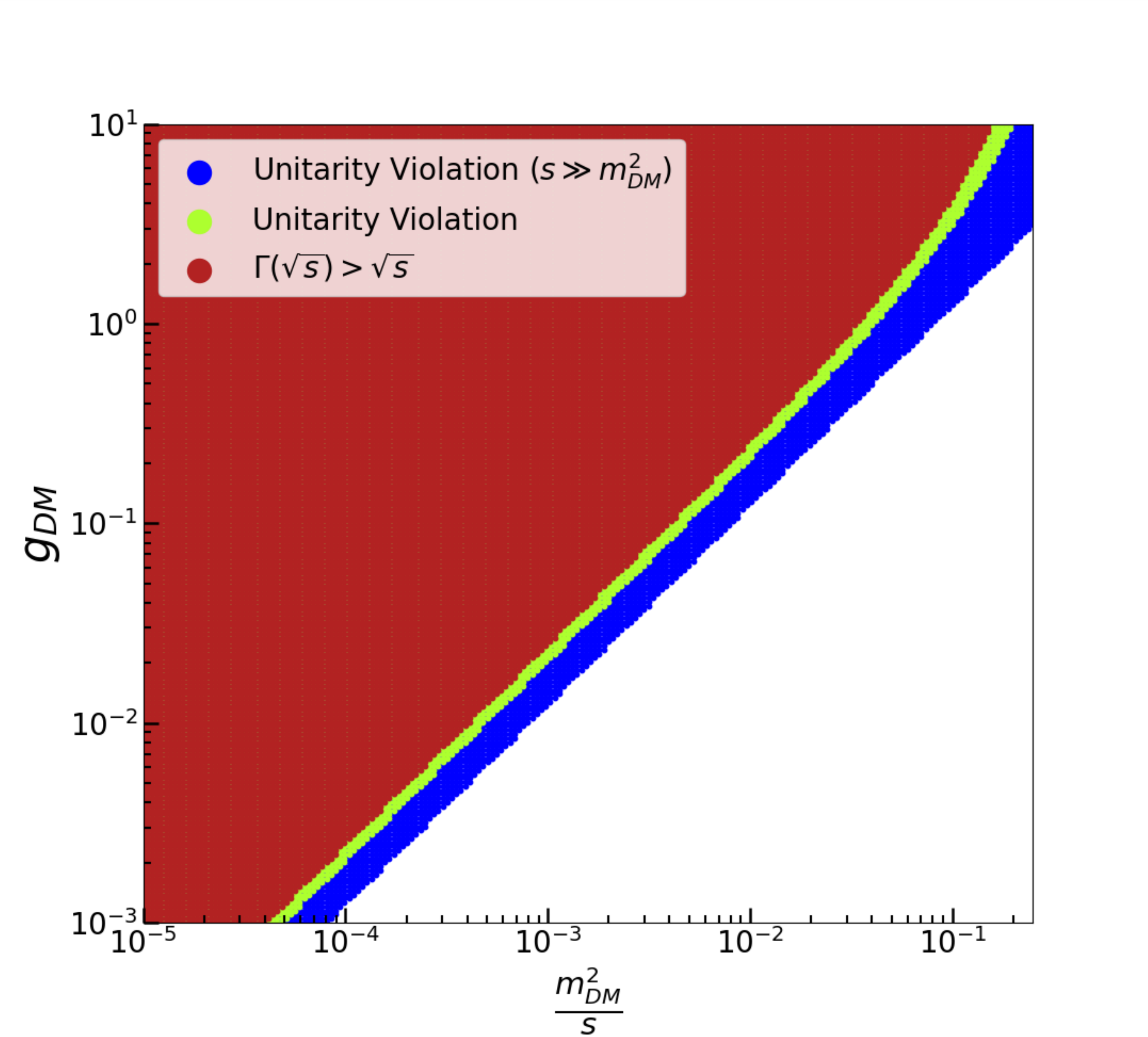}
  \caption{Comparison between unitarity violation and unphysical decay widths for a demonstrative choice of parameters ($s = 10^{8} \, \mathrm{GeV}^2$, $m_{\text{M}} = \sqrt{s}$, $g_{\text{q}}=0$, and varying $m_{\text{DM}}$ and $g_{\text{DM}}$). The requirement of a physical off-shell decay width (red) excludes a smaller region than the requirement of unitarity (green), but follows a similar trend. Taking the high-energy limit of the unitarity bound is a consistently stricter cut on the parameter space (blue). }
  \label{ComparisonDecayWidthUnitarity}
\end{figure}

Alongside unitarity violation, another indication that the model breaks down is that the decay width of the mediator becomes large, indicating the inapplicability of perturbation theory to the underlying scattering process. When the mediator is on-shell, this can be interpreted as a bound on the decay width
\begin{align}
\mathrm{\Gamma} (m_{\text{M}}) \leq m_{\text{M}}\,.
\label{eq:width}
\end{align}
We reject all points in parameter space that do not satisfy this bound. In the following we require that an analogous inequality also holds for the off-shell decay width when replacing $m_{\text{M}}$ by $\sqrt{s}$:
\begin{align}
\mathrm{\Gamma} (\sqrt{s}) \leq \sqrt{s}\,.
\end{align}
In the high energy limit, the bound on the off-shell decay width results in the requirement
\begin{align}
   s \leq  \frac{\sqrt{192 \pi} m_{\text{DM}}^2}{g_{\text{DM}}}\,.
\end{align}
This differs from eq.~\eqref{UnitatityBoundEq} by a factor of 2 (the unitarity bound being the stricter of the two). When assuming high collision energies, it is therefore clear to see that the unitarity bound and off-shell decay width bound are practically interchangeable. Figure~\ref{ComparisonDecayWidthUnitarity} shows a comparison between the unitarity constraint with and without taking the high-energy limit, for a representative choice of parameters, along with the exclusion from requiring that the off-shell decay width is physical. The similarity between the unitarity and decay width conditions would suggest that for the choice of parameters shown, very little difference would be observed if the two were interchanged. 

\section{Constraints}
\label{Constraints}

Interactions between DM and SM quarks are constrained by many different measurements of astrophysical, cosmological and particle physics processes.

We use likelihoods, implemented in \GB \textsf{2.4}, for DM direct and indirect detection experiments, collider searches at the ATLAS and CMS experiments, and the measurement of the DM relic abundance. We generate the necessary model-specific \GB module functions (including those used to store spectrum and decay information \cite{SDPBit}) using the GAMBIT Universal Model Machine (\gum)~\cite{GUM}. This includes interfaces to backend codes that contain physics calculations for each DM observable. We apply the perturbative unitarity and physical off-shell decay width constraints described in section~\ref{sec:UnitarityBound} to the calculation of collider signals, to ensure that calculations are accurate and the resulting limits are conservative; this is detailed in section~\ref{constraints:monojet}. We reject parameter points that fail the requirement of a physical on-shell decay width of the mediator, before calculating their likelihood contributions.

Table \ref{tab:experiments} provides a summary of each likelihood that we include that is sensitive to BSM physics. For each likelihood, we provide either: $\ln \mathcal {L}^\text{bg}$, the value that the likelihood takes purely from the SM, or $\ln \mathcal {L}^\text{max}$, the best-case likelihood that can be achieved when parameters exactly match their centrally measured values.

For a detailed description of the implementation of each likelihood in \GB, we refer the reader to the previous work in this series~\cite{DMSimpI}. We provide brief summaries of each likelihood in the following subsections.

\begin{table}[t]
  \centering
  \begin{tabular}{lcc}
    \toprule
    \textbf{Experiment} & \textbf{$\ln \mathcal {L}^\text{bg}$} & \textbf{$\ln \mathcal {L}^\text{max}$}    \\ \midrule
        CDMSlite~\cite{Agnese:2015nto} & $-16.68$ \\
        CRESST-II~\cite{Angloher:2015ewa} & $-27.59$ \\
        CRESST-III~\cite{Abdelhameed:2019hmk} & $-27.22$ \\
        DarkSide 50~\cite{Agnes:2018fwg} & $-0.09$ \\
        LUX 2016~\cite{LUXrun2} & $-1.47$ \\
        PICO-60~\cite{Amole:2017dex,Amole:2019fdf} & $-1.496$ \\
        PandaX~\cite{Tan:2016zwf,Cui:2017nnn,PandaX-4T:2021bab} & $-6.121$ \\
        XENON1T~\cite{Aprile:2018dbl} & $-3.651$ \\
        LZ 2022~\cite{LZ:2022ufs} & $-4.636$
        \\[2mm]
        LHC Dijets~\cite{CMS:2019gwf,ATLAS:2019fgd,ATLAS:2018qto,CDF:2008ieg,ATLAS:2018hbc,ATLAS:2018hzj,CMS:2019emo,ATLAS:2019itm,CMS:2019xai} & $0$ \\
        ATLAS monojet~\cite{Aad:2021egl} & $0$ \\
        CMS monojet~\cite{CMS:2021snz} & $0$
        \\[2mm]
        \emph{Fermi}-LAT~\cite{LATdwarfP8} & $-33.245$ \\
        \emph{Planck} 2018: $\Omega h^2$~\cite{Aghanim:2018eyx} & & $5.989$
        \\[2mm]
        Nuisances (see Table~\ref{tab:parameters}) & & $-5.995$
        \\[1mm] \bottomrule
  \end{tabular}
  \caption{All likelihoods included in our fits. We give the SM-only (i.e.\ background-only) log-likelihood $\ln \mathcal {L}^\text{bg}$ for those that search for events above an SM background. For the rest, we give the highest achievable value of the log-likelihood $\ln \mathcal {L}^\text{max}$, where the predicted value of the chosen observable or a nuisance parameter is exactly equal to its measured value.}
  \label{tab:experiments}
\end{table}

\subsection{Relic Density}

We use \gum to generate the \CH \textsf{v3.6.27} \cite{Pukhov:2004ca,Belyaev:2012qa} model files that are supplied to \mo \textsf{v3.6.9.2}\cite{micromegas}. The relic density of DM is obtained with the \darkbit interface which uses \mo to solve the Boltzmann equation for the number density of DM particles in thermal equilibrium, assuming a standard cosmological history. To form a likelihood from the relic abundance, we compare the calculated density to the \emph{Planck} 2018 measurement of $\Omega_{\textrm{DM,obs}}\,h^2 = 0.120 \pm 0.001$~\cite{Aghanim:2018eyx} with a $1\%$ theoretical error added in quadrature with the quoted \emph{Planck} uncertainty.

We study both cases where the DM candidate is a subcomponent of the observed relic abundance and where it fully saturates the abundance. When requiring that it saturates the relic abundance, we use the Planck measurement to define a Gaussian likelihood based on the predicted WIMP abundance. When allowing it to form a subcomponent, we modify this likelihood to be flat for predicted densities below the measurement; details can be found in Ref.\ \cite{gambit}.

\subsection{Direct Detection}

The parameters of a simplified DM model can be translated to the coefficients of the relevant operators in a non-relativistic EFT for WIMP-nucleon scattering, $c_{\mathrm{i}}^{\mathrm{N}}(q^2)$. The single relevant operator and its coefficient for the vector DM simplified model in this study is~\cite{Baum:2017kfa}.
\begin{align}
  c_1^{\mathrm{N}} = - \frac{2 g_{\text{q}} g^{\text{V}}_{\text{DM}}}{m_{\text{M}}^2}\,,
\end{align}
which was supplied to \textsf{DDCalc v2.2.0}~\cite{DarkBit,HP}, to compute the differential cross-section and target element of interest. We do not include the effect of operator mixing from running as it has been shown to have little effect for pure vector couplings of the mediator to quarks~\cite{DEramo:2016gos}.

We calculate direct detection likelihoods from the most recent XENON1T analysis~\cite{Aprile:2018dbl}, LUX 2016~\cite{LUXrun2}, PandaX 2016, 2017 and 4T~\cite{Tan:2016zwf,Cui:2017nnn,PandaX-4T:2021bab}, CDMSlite~\cite{Agnese:2015nto}, CRESST-II and CRESST-III~\cite{Angloher:2015ewa, Abdelhameed:2019hmk}, PICO-60 2017 and 2019~\cite{Amole:2017dex,Amole:2019fdf}, DarkSide-50 \cite{Agnes:2018fwg} and LZ~\cite{LZ:2022ufs}\footnote{The description of how LZ is implemented is provided in Ref.~\cite{DMSimpI}.}.

\subsection{Indirect Detection}

The model we study has two primary DM annihilation channels, annihilation to mediators and to quarks. Annihilation to a pair of mediators occurs as an $s$-wave process, and will be the primary annihilation channel when kinematically allowed ($m_{\text{DM}} > m_{\text{M}}$). When this channel is closed, the annihilation will occur to a pair of quarks, through the suppressed $p$-wave channel. We do not include $p$-wave contributions to the gamma-ray flux as they should not be large enough to impact searches toward dwarf spheroidals for the model we consider.

We compute the annihilation cross-section with \CH, using the \gum interface to generate the required \CH model files. We use the combined analysis of 15 dwarf spheroidal galaxies, \texttt{Pass-8}, performed by the \emph{Fermi}-LAT Collaboration over 6 years of data taking~\cite{LATdwarfP8}, using \gamlike~\textsf{v1.0.1} to compute the likelihood through its interface to \darkbit. DM annihilations at the centre of our galaxy are an alternative to dwarf spheroidal measurements. Since \emph{Fermi}-LAT Galactic Centre limits are not as robust as limits from dwarf spheroidals, we do not include them in this study. We do however briefly comment on the future impact of CTA observations on the parameter space of this model in section \ref{section:discussion}.

\subsection{Monojet searches at the LHC}
\label{constraints:monojet}

One of the primary channels via which to search for the model at colliders is the creation of a pair of final state WIMPs in association with a jet created by initial state radiation. This gives a signature of a single jet plus missing transverse energy ($\slashed{E}_{\mathrm{T}}$). We include the most current monojet searches from CMS and ATLAS searches with $137\,\rm{fb}^{-1}$ \cite{CMS:2021snz} and $139\,\rm{fb}^{-1}$ \cite{Aad:2021egl} of Run II integrated luminosity respectively.

To calculate the total production cross-section $\sigma$ and the product of the efficiency and acceptance for passing the analysis kinematic selections $\epsilon A$, we perform simulation of Monte Carlo events with \textsf{MadGraph\_aMC@NLO}~\cite{Alwall:2011uj} (\textsf{v3.1.1}), interfaced to \pythia \textsf{v8.3} \cite{Sjostrand:2007gs} for parton showering and hadronization. To form the quantity $\epsilon A$ we pass these events through \madanalysis 5 \cite{Conte:2012fm} and implement the ATLAS and CMS monojet analyses. Rather than perform this calculation for each parameter sample, we precompute a grid of cross sections (\(\sigma\)) and $\epsilon A$ factors in advance, and interpolate them at runtime using \colliderbit \cite{colliderbit}.

An additional analysis cut is added to our implementations of the ATLAS and CMS kinematic selections, to remove any events which would violate the unitarity bound presented in section \ref{Unitarity Violation}, replacing $\sqrt{s}$ with the invariant mass of the DM pair. When this cut becomes strong enough, there is a significant drop in the predicted acceptance of the analysis, and we can no longer make any sensible predictions regarding collider constraints. If no simulated events pass the unitarity cut, we expect the parameter point to be unobservable at the LHC and simply assign the background-only likelihood.

The interpolation grid we use is as follows:
\begin{itemize}
\item \textbf{mediator mass}: 17 values, 50\,GeV--10\,TeV
\item \textbf{DM/mediator mass ratio}: 16 values, 0.01--50
\item \textbf{quark-mediator coupling}: 5 values, 0.01--1.0
\item \textbf{DM-mediator coupling}: 7 values, 0.01--3.0
\end{itemize}

The grids for the mediator mass and couplings were chosen to be approximately equally spaced in log-space. The ratio of DM and mediator masses is more effective than the DM mass as a grid variable as it allows us to choose a grid with a higher density of points across the resonance region, where we expect rapid changes in predictions. Below the DM mass/mediator mass ratio of 0.01, we assume that we can safely extrapolate to small DM masses as the predicted signal should not vary significantly. After removing any points with DM masses above the limits of our scan, this gives a total number of 6370 grid points.

\subsection{Searches for dijet resonances}

The presence of a mediating particle in the model may generate dijet events at colliders,  with an invariant mass of approximately the mediator mass. Dijet resonance searches provide robust constraints on DM simplified models, where the extremely high multijet background must be removed with clever kinematic analysis cuts.

The cross-section for the production of a dijet resonance can be approximated as the product of the cross-section of mediator production and the branching ratio of the mediator into quarks, assuming that the narrow width approximation holds. When the ratio of the mediator decay width to mass is high, this approximation breaks down, and our treatment of dijet searches would become dubious. We briefly investigate the dependence of the model exclusion on this assumption in section~\ref{Results}.

We implement dijet limits provided by ATLAS and CMS~\cite{CMS:2019gwf,ATLAS:2019fgd,ATLAS:2018qto,CDF:2008ieg,ATLAS:2018hbc,ATLAS:2018hzj,CMS:2019emo,ATLAS:2019itm,CMS:2019xai} by scaling of the published limits of the mediator-quark coupling by the branching ratio into quarks, following the same approach as Refs.~\cite{Bagnaschi:2019djj,DMSimpI}. These published limits are interpolated in $m_\mathrm{M}$ for each parameter point, and the likelihood is formed from the most constraining search for a given mediator mass. The combined coupling upper limits are provided in Figure 1 of Ref.~\cite{DMSimpI}.

In the absence of tree-level couplings of the mediator to leptons, couplings at loop level may still be generated through kinetic mixing, and the model may be observable at dilepton searches. Despite the tight constraints on dilepton signatures for vector mediated simplified models, the loop suppression of these lepton couplings will prevent dilepton constraints on the quark coupling being any stronger than dijet limits. For this reason we do not include dilepton constraints in this study. For a discussion on the lepton couplings generated through kinetic mixing, we refer the reader to Refs.~\cite{Duerr:2016tmh, Kahlhoefer:2015bea}.

\begin{table}
\centering
\begin{tabular}{ll}
\toprule
\textbf{Parameters}          & \textbf{Range} \\ \midrule
DM mass, \(m_\mathrm{DM}\)          & \([50,10000]$\,GeV     \\ \hline
Mediator mass, \(m_\mathrm{M}\)     & \([50,10000]$\,GeV    \\ \hline
quark-mediator coupling, \(g_\mathrm{q}\)     & \([0.01,1.0]\)    \\ \hline
mediator-DM coupling (vector), \(g^{\rm V}_\mathrm{DM}\)   & \([0.01,3.0]\) \\ \hline
\textbf{Nuisance Parameters} & \textbf{Value} ($\pm 3 \sigma$ range) \\ \midrule
Local DM density, \(\rho_0\)   & \([0.2,0.8]$\,GeV\,cm$^{-3}\)  \\ \hline
Most probable DM speed, \(v_\mathrm{peak}\) & \(240 (24)$\,km\,s$^{-1}\)     \\ \hline
Galactic escape speed, \(v_\mathrm{esc}\)   & \(528 (75)$\,km\,s$^{-1}\) \\
\bottomrule
\end{tabular}
\caption{List of model and nuisance parameters and their corresponding scan ranges.}
\label{tab:parameters}
\end{table}

\subsection{Nuisance Parameter Likelihoods}
\label{sec:nuisances}

Along with the model parameters in the model we study, we also include a set of nuisance parameters which are used in each of our astrophysical likelihoods. A complete list of these parameters is given in Table~\ref{tab:parameters}.

\begin{table*}[t]
\centering
\begin{tabular}{lccccc}
\toprule
Relic Density & Best Fit \(m_\mathrm{DM}\) (GeV) & Best Fit \(m_\mathrm{M}\) (GeV) & Best Fit \(g_\mathrm{q}\) & Best Fit \(g^{\rm V}_\mathrm{DM}\)  & \(\Delta \ln \mathcal{L}\) \\ \midrule
Upper limit  & 4950 & 9960 & 0.010 & 1.041  & 0.00  \\ \hline
All DM  & 4570 & 9210 & 0.016 & 0.763  & -0.45 \\ \bottomrule
\end{tabular}
\caption{Approximate best-fit points for each scan. $\Delta \ln \mathcal{L}$ values are defined as $\ln \mathcal{L} - \ln \mathcal{L}^{\mathrm{ideal}}$, where the ideal likelihood is the combination of background-only and maximum possible likelihoods detailed in Table \ref{tab:experiments}.}
\label{tab:BestFitPoints}
\end{table*}

\begin{figure*}[tbp]  
  \centering
  \includegraphics[width=\columnwidth]{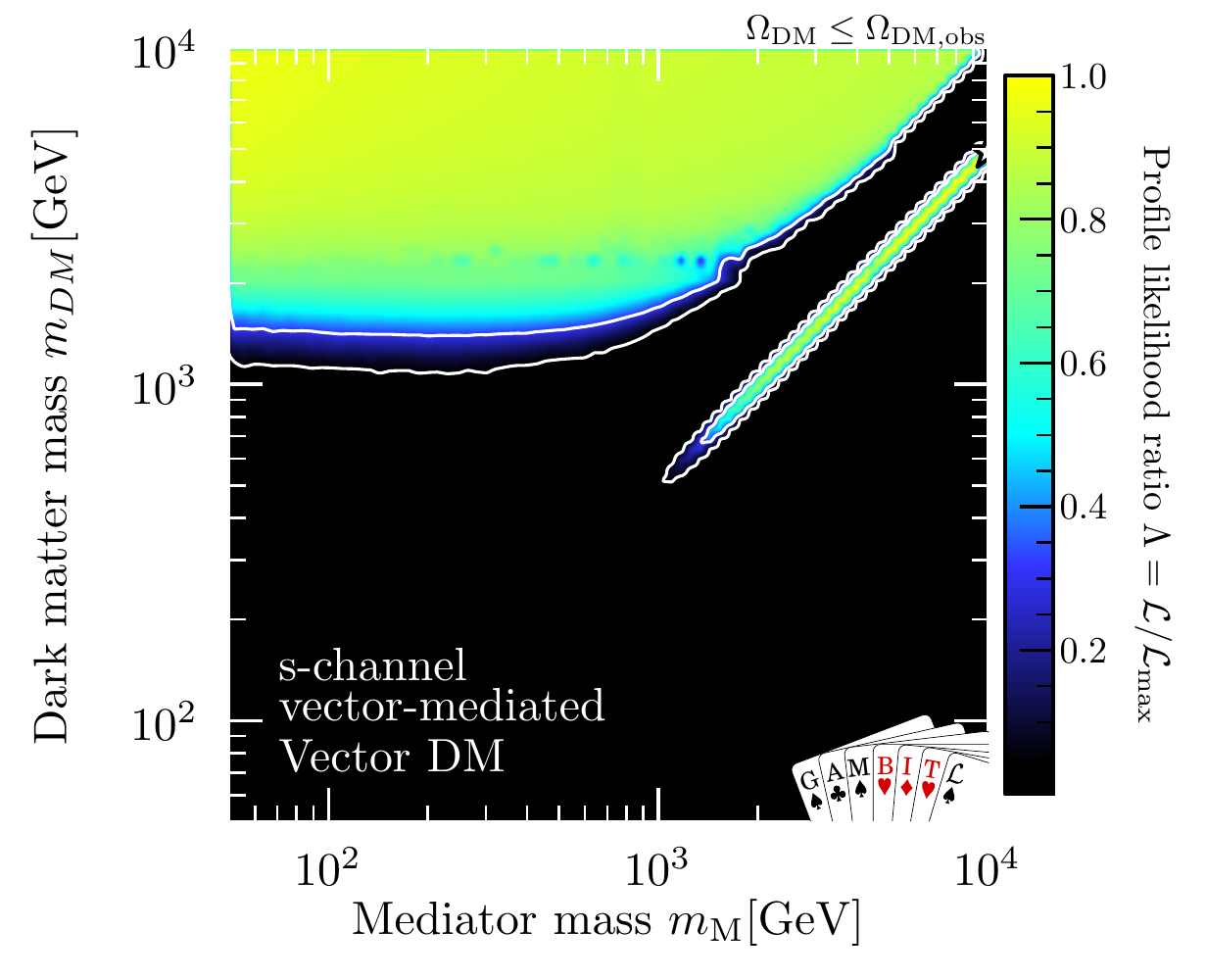}
  \includegraphics[width=\columnwidth]{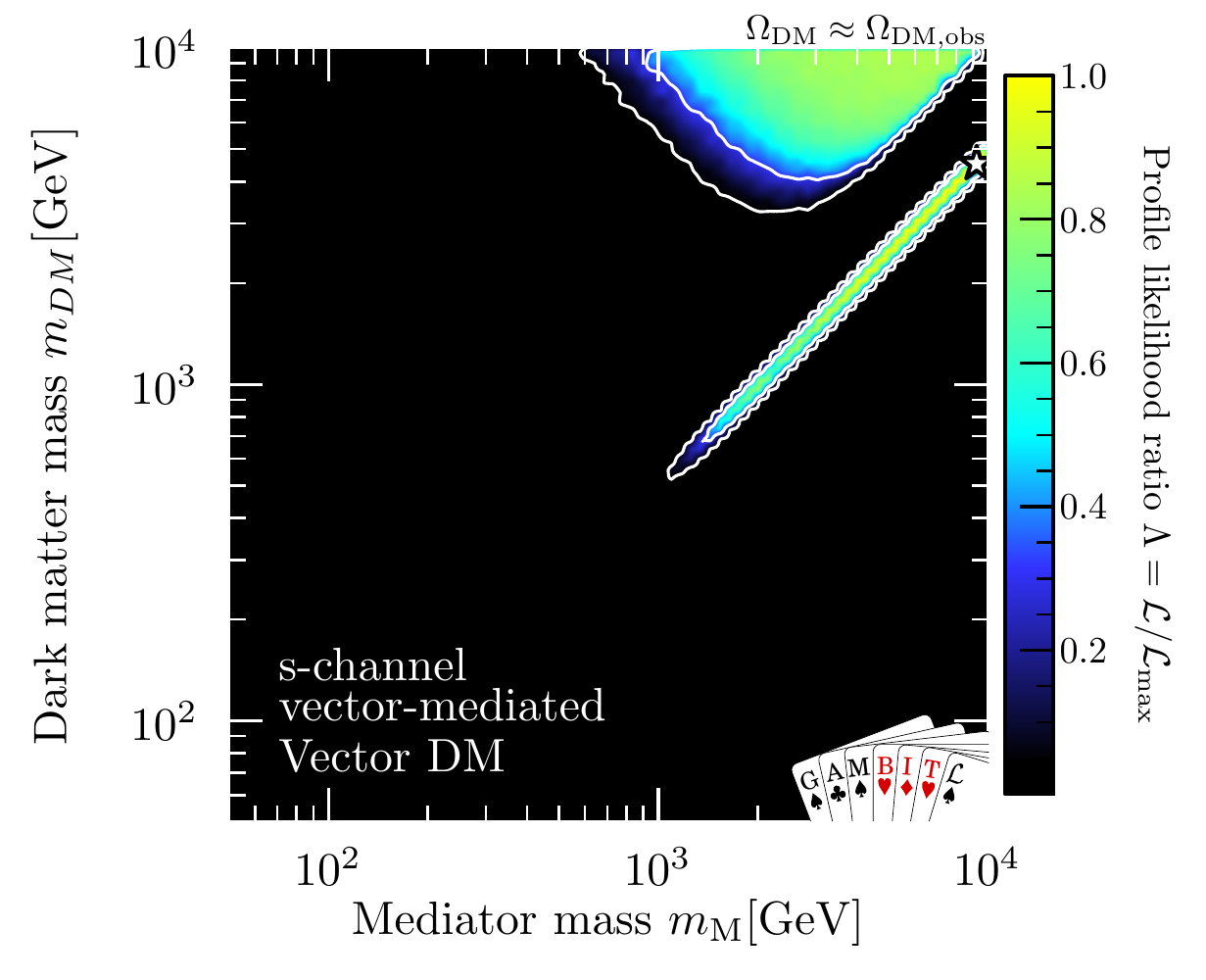}
  \caption{Profile likelihood, profiled over couplings. The measured DM relic abundance is taken as an upper limit (left) or to be composed entirely of the vector DM candidate (right). 1$\sigma$ and 2$\sigma$ contours are shown in white, with the star representing the best-fit point.}
  \label{ProfileLike}
\end{figure*}

\begin{figure*}[tbp]  
  \centering
  \includegraphics[width=\columnwidth]{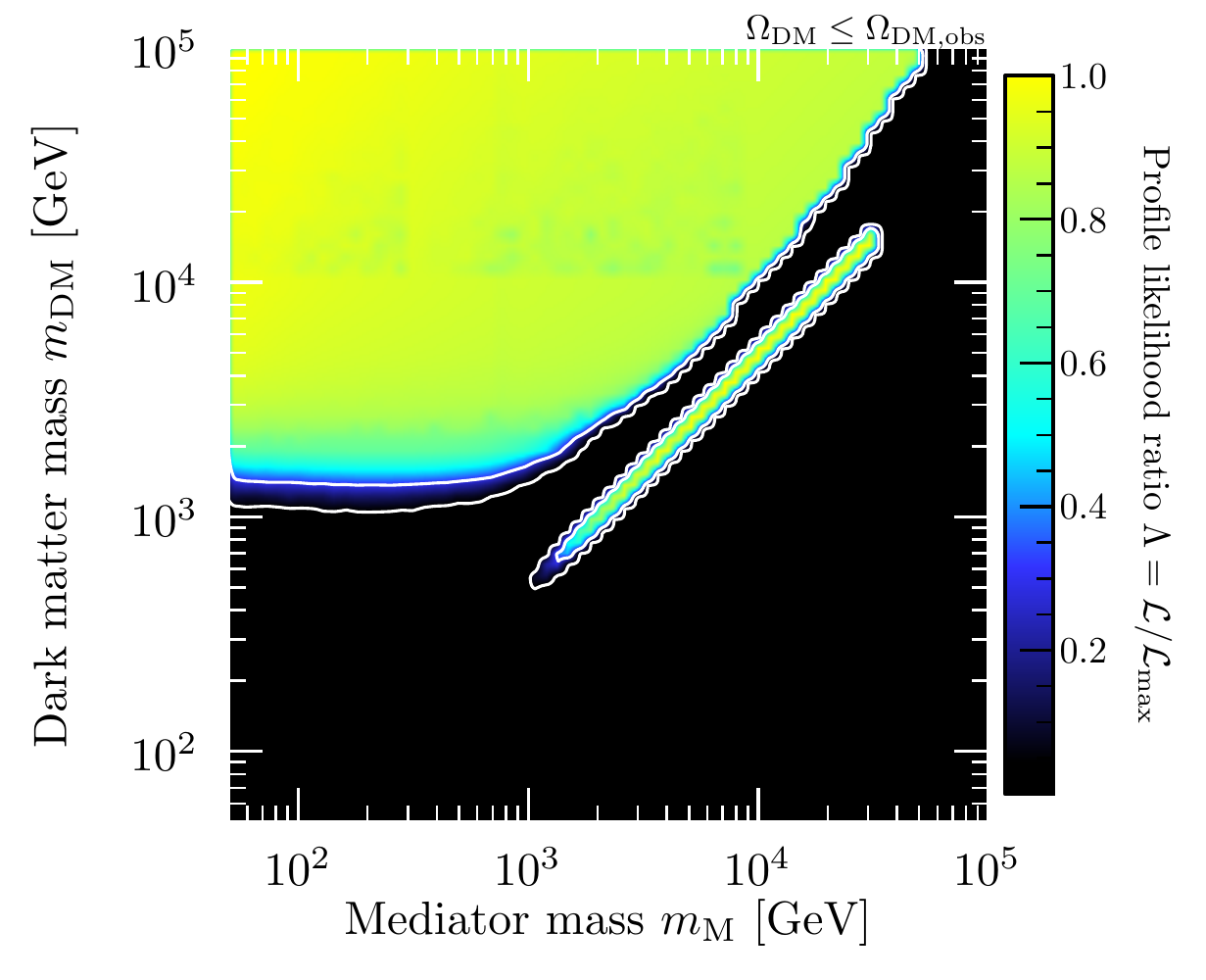}
  \caption{Profile likelihood, profiled over couplings, for DM and mediator masses up to $10^5$ GeV. The measured DM relic abundance is taken as an upper limit. 1$\sigma$ and 2$\sigma$ contours are shown in white.}
  \label{ProfileLike_Extended}
\end{figure*}

We treat the local DM density $\rho_0$ following the standard procedure in \darkbit, where $\rho_0$ is assumed to be log-normally distributed, centred around $\rho_0 = 0.40$\,GeV\,cm$^{-3}$ and with an error $\sigma_{\rho_0}=0.15$\,GeV\,cm$^{-3}$. The scan range of $\rho_0$ is asymmetric to reflect this distribution. $3 \sigma$ ranges for all other nuisance parameters are provided in Table~\ref{tab:parameters}.

We treat the Milky Way halo in the same way as in several of our previous DM studies~\cite{HP,DMEFT,DMSimpI}, where the DM velocity is assumed to follow a Maxwell-Boltzmann distribution. The peak velocity and Galactic escape velocity uncertainties are described by Gaussian likelihoods with $v_{\rm{peak}} = 240\, \pm \,8$\,km\,s$^{-1}$ \cite{Reid:2014boa} and $v_{\rm{esc}} = 528 \pm 25$\,km\,s$^{-1}$ (based on \emph{Gaia} data \cite{Deason:2019kgj}), respectively.


\section{Results}
\label{Results}

We have performed a comprehensive scan of the model parameter space using the differential evolution sampler \textsf{Diver v1.0.4} \cite{ScannerBit} with a convergence threshold of $10^{-6}$ and a population of $20\,000$, with an additional scan for DM masses below 2 TeV to improve sampling. We carried out two separate scans for the case where the observed DM relic density is taken as an upper limit or as a two-sided measurement. Unlike the previous study in this series~\cite{DMSimpI}, scans with a capped LHC likelihood were not performed, as any small preferences over the background-only hypothesis in mono-jet searches were not found to occur within the surviving parameter space of the scan. A scan with a capped LHC likelihood would therefore produce results that were indistinguishable from its uncapped equivalent. 

Table~\ref{tab:parameters} provides the full list of parameters and scan ranges. We adopt the same choice of scan ranges and sampling distributions of the masses and couplings as those in Ref.~\cite{DMSimpI}. Very small couplings are avoided in order to focus on regions where unitarity violation may be relevant. The coupling upper bounds are of order unity in order to keep the decay width of the mediator from becoming excessively large. The range of masses was chosen to focus on regions where it was expected that both collider searches and direct and indirect searches may be complementary. The parameter points that give the best likelihoods are given in Table~\ref{tab:BestFitPoints}.

\begin{figure*}[tbp]
  \centering
  \includegraphics[width=\columnwidth]{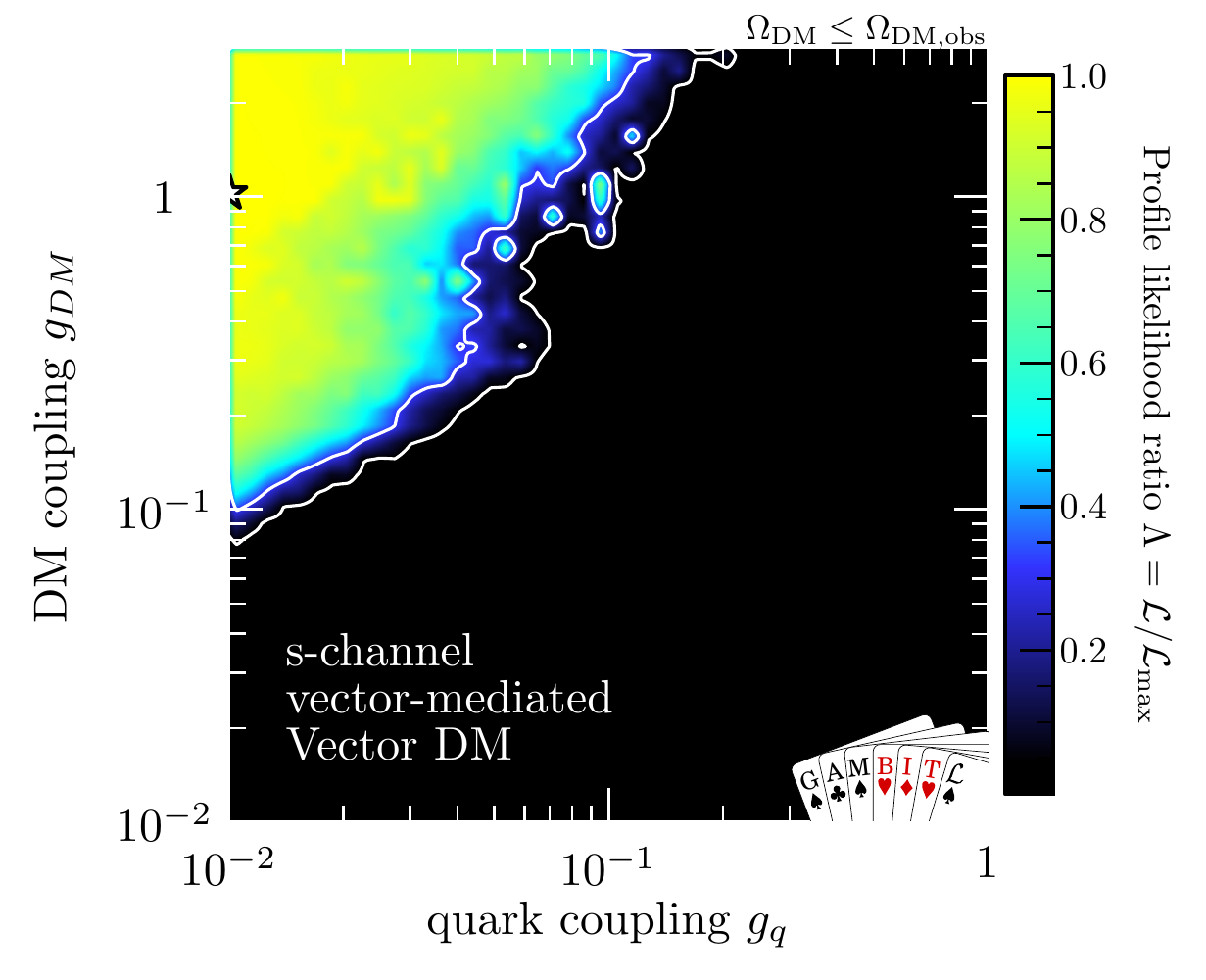}
  \includegraphics[width=\columnwidth]{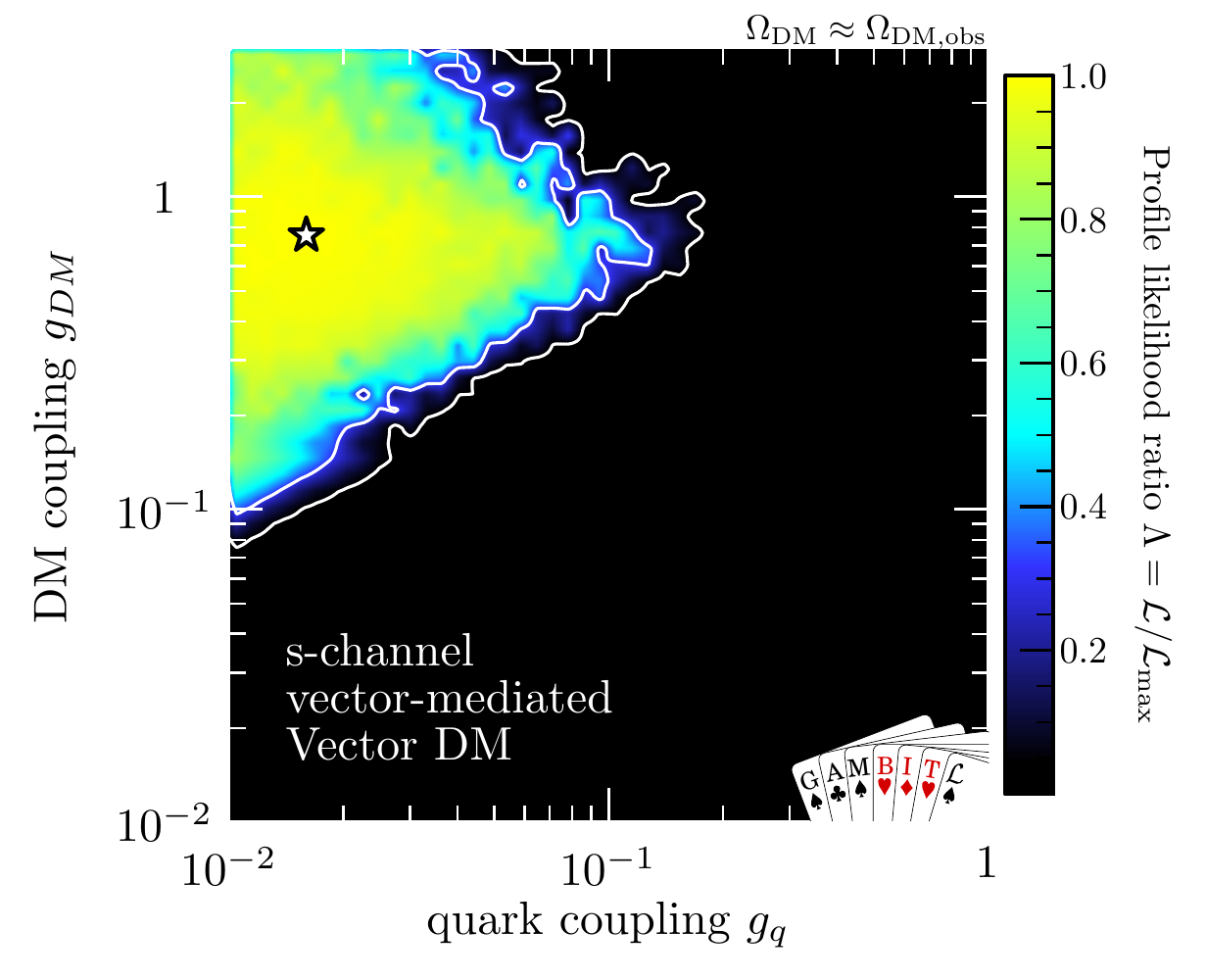}
  \caption{Profile likelihood, profiling over mediator and DM masses, for a relic abundance upper limit (top) and a saturated relic abundance (bottom).  1$\sigma$ and 2$\sigma$ contours are shown in white, with the star representing the best-fit point.}
  \label{ProfileLike_couplings}
\end{figure*}

\begin{figure*}[tbp]
  \centering
  \includegraphics[width=\columnwidth]{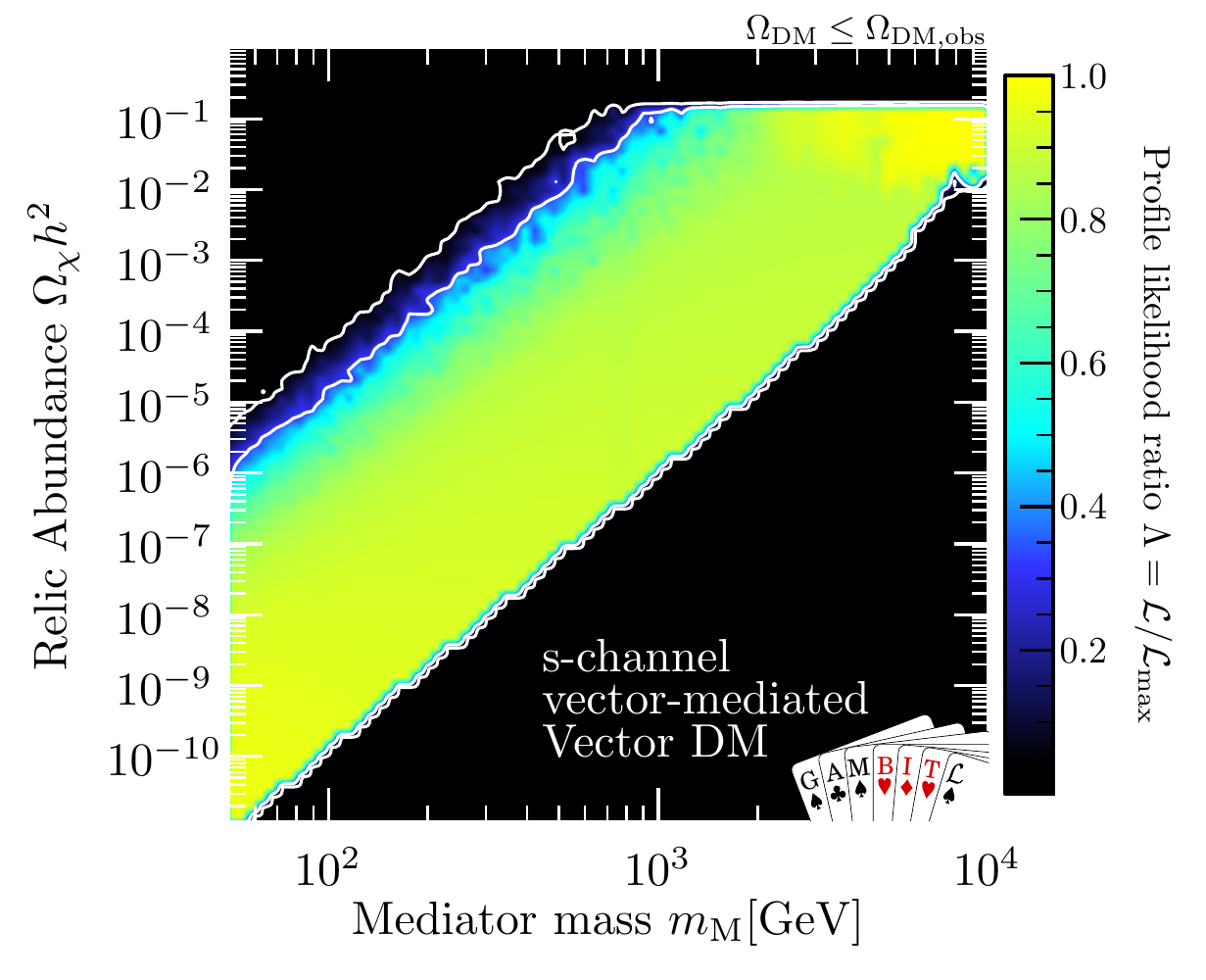}
  \includegraphics[width=\columnwidth]{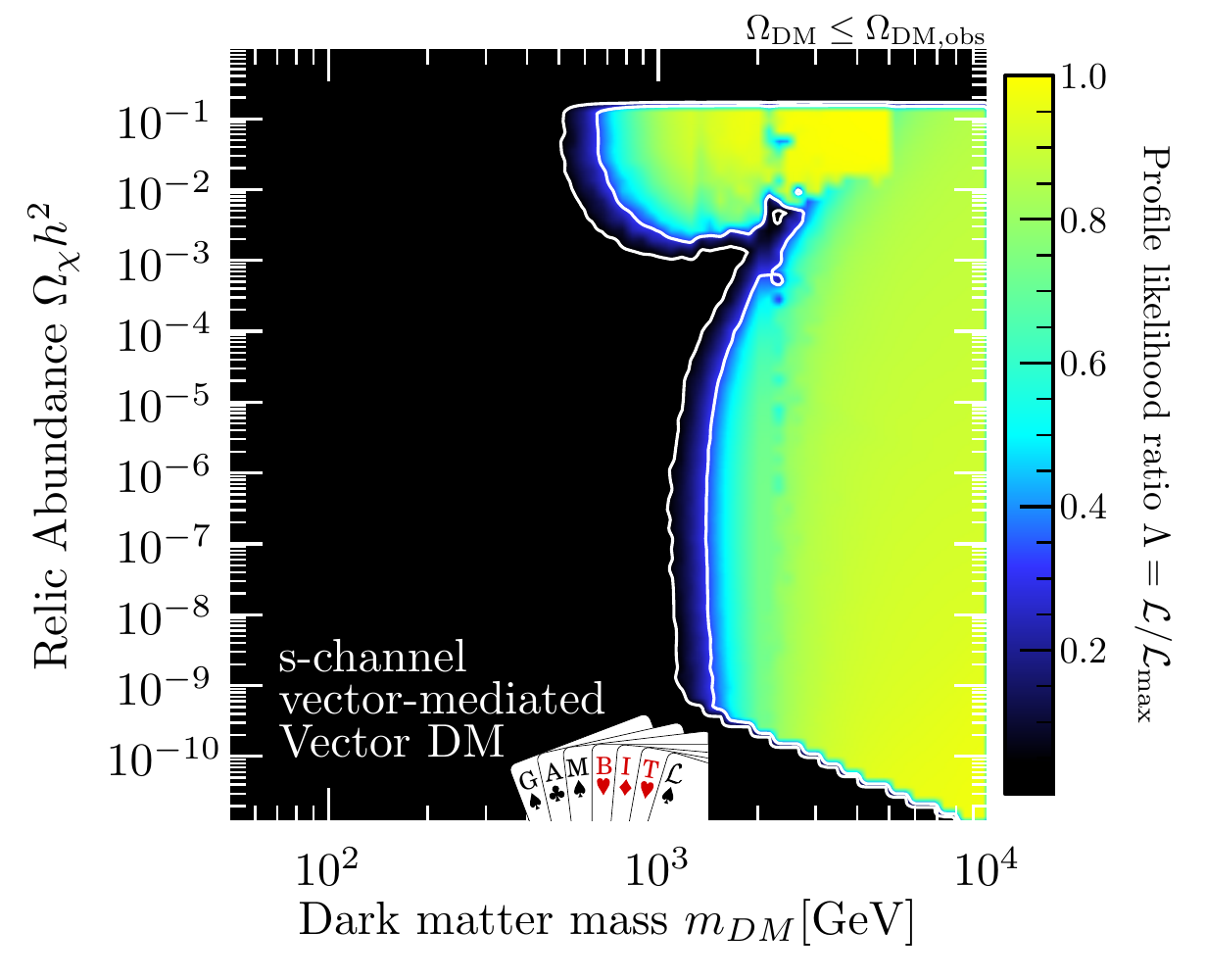}
  \caption{DM relic abundance for the surviving parameter space, when taking the relic abundance measurement as an upper limit.  We show the abundance both against mediator mass (left) and against DM mass (right).  1$\sigma$ and 2$\sigma$ contours are shown in white.  }
  \label{oh2}
\end{figure*}

\begin{figure}[tbp]
  \centering
  \includegraphics[width=\columnwidth]{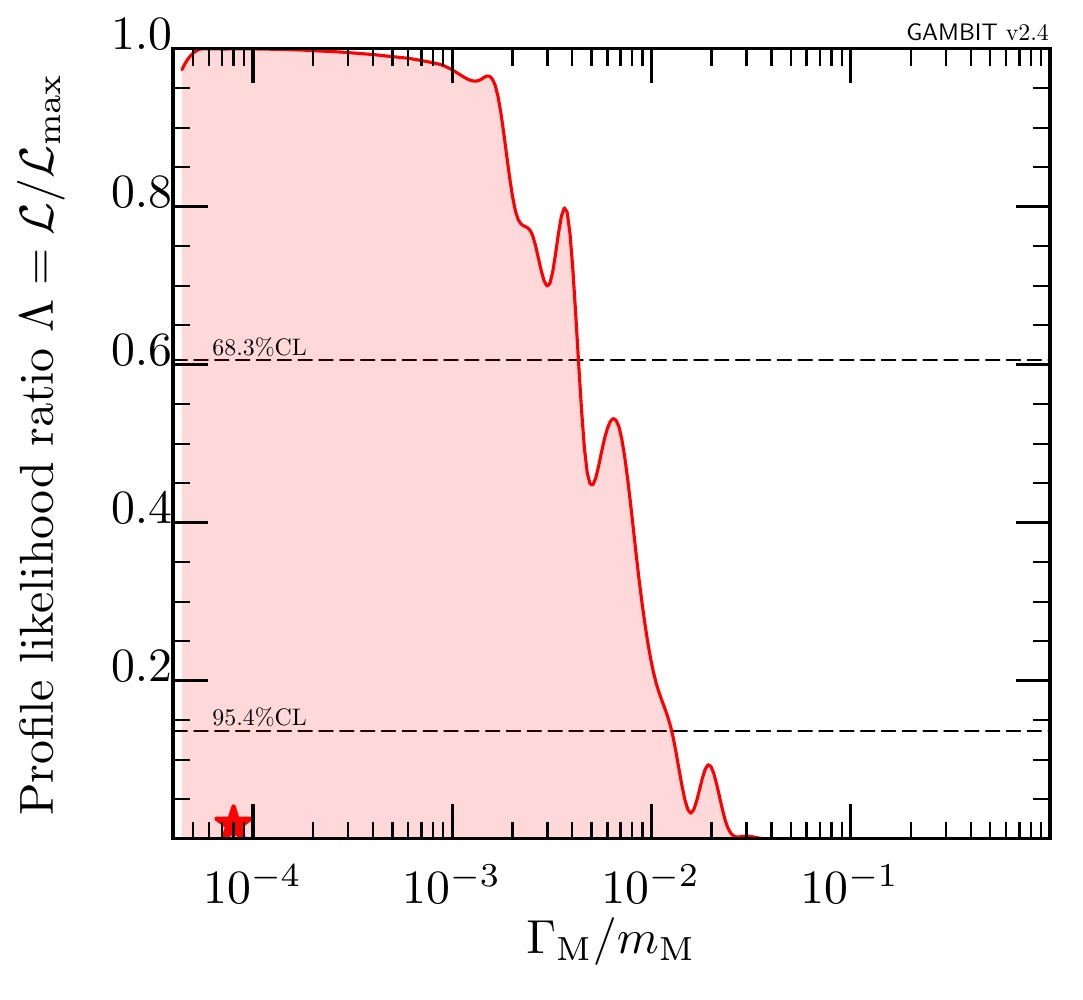}
  \caption{Profile likelihood, as a function of the mediator width to mass ratio, profiled over all model parameters. 1$\sigma$ and 2$\sigma$ confidence limits are shown in black, with the red star representing the best-fit point.}
  \label{NWARatio}
\end{figure}

The profile likelihood from combined constraints on the complex vector DM model is shown in Figures~\ref{ProfileLike} and~\ref{ProfileLike_couplings}. The model prefers parameter regions where DM annihilation is efficient, and there are two regions corresponding to the two DM annihilation channels. Around the diagonal $m_{\text{M}} \approx 2 m_{\text{DM}}$, the annihilation occurs close to a resonance into a pair of quarks. For regions where $m_{\text{DM}} > m_{\text{M}}$, the annihilation occurs as a $t$-channel process into a pair of mediator particles. Below approximately 500 GeV, the annihilation may not be great enough to prevent exclusion from direct detection constraints without leaving the limits of the scanned parameter ranges. 

This shape is highly similar to those presented for a scalar DM candidate in Ref.~\cite{DMSimpI}. This is because the strongest limits come from the direct detection experiments, which are dependent on the effective operators that are relevant, and this model shares the same relevant operator as the scalar DM model. The model survives for a greater proportion of the parameter space than the scalar DM model, despite the additional inclusion of PandaX-4T direct detection data in this work. The small variation in the profile likelihood around 2 TeV is a sampling artifact, and does not reflect any physical change in predictions. 

In Figure~\ref{ProfileLike_Extended}, we show how the profile likelihood changes if the scan range were extended to masses up to $100$ TeV. The resonance region closes off around 30 -- 40 TeV as DM becomes overabundant unless the couplings become non-perturbative. The non-resonant region continues on with a largely flat likelihood. For DM masses well beyond 100 TeV, thermally produced DM will violate generic unitarity bounds~\cite{Griest:1989wd}.

Requiring that the DM relic abundance is saturated shrinks the surviving region to mediator masses above 1 TeV for the off-resonance region. For lower mediator masses, the non-relativistic effective coupling to nucleons is stronger and therefore expected signal at direct detection experiments is greater. Figure~\ref{oh2} (left) shows that at low mediator masses, the likelihood is higher in parameter regions where the model strongly underproduces DM to avoid tension with these experiments. As the strength of the direct detection constraints increases toward lower DM mass, the surviving parameter region also has a lower bound on the DM mass that may be seen in Figure~\ref{oh2} (right). The surviving region along the resonance does not depend strongly on whether the abundance likelihood is taken as a one-sided upper limit or a two-sided measurement. Measurements of dwarf spheroidal galaxies do not appear to have any strong influence on the profile likelihoods.

We find that, in the surviving parameter regions, the decay width of the mediator is dominated by the partial width to quarks. Limits from dijet searches prevent mediator-quark couplings $g_{\text{q}}$ above roughly 0.1 for most of the parameter space. This preference toward lower $g_{\text{q}}$ reduces the effect of high decay widths, as the partial width to quarks is proportional to $g_{\text{q}}^2$. Fig~\ref{NWARatio} shows that within $2 \sigma$ of the best-fit point, the decay width of the mediator does not exceed $0.02 m_{\text{M}}$, safely satisfying the narrow width requirement.

The effect of monojet searches cannot be seen directly on the results of the profile likelihood. For any model parameters where monojet searches would have sensitivity, these are strongly excluded by relic abundance limits and direct detection searches. The combined global fit therefore does not appear to be strongly affected by unitarity considerations. This conclusion might however change when considering a more general parameter space including also the couplings $b_6$ and $b_7$.

The best fit for each scan lies along the resonance, at the upper limits of the masses, and toward the lower limits of the quark coupling. In these regions, the relic abundance, and the strength of the direct detection signals are minimised. When the DM candidate is allowed to be a subcomponent of the observed DM density, this best fit point approximately matches the background likelihood as the signals at any given DM experiment are almost entirely negligible. We compute an approximate $p$-value of the best-fit likelihood conditioned on the `ideal' scenario (sum of background-only and max entries in Table~\ref{tab:experiments}) for 1--2 effective degrees of freedom. Further explanation of the construction of this particular $p$-value can be found in Ref.~\cite{SSDM}. Neither case (saturated or subdominant DM) is disfavoured, returning $p$-values of 0.3 and above.

We limited the couplings to be no lower than 0.01, in order to target parameter regions where unitarity violation was most likely to cause issues without introducing large hierarchies between couplings. If the scan range was expanded to include smaller $g_{\text{q}}$, it can be seen from Figure~\ref{ProfileLike_couplings} how the size of the surviving parameter space should increase. Expanding the lower limit on $g_{\text{DM}}$ will only expand the surviving space if $g_{\text{q}}$ is also expanded. For the parameters scanned over in this work, the model is excluded for lower $g_{\text{DM}}$, as there cannot be sufficient annihilation of the thermal DM abundance.

\section{Discussion}
\label{section:discussion}

In this work, we have derived a unitarity bound for a simplified model with a vector DM candidate that interacts with SM quarks via an $s$-channel vector mediator. We showed that this unitarity bound is highly similar to the bound on the model parameters one would require from the behaviour of the off-shell decay width, which is another challenge that plagues these theories. Applying this bound to simulated collider events, we performed a global scan of this model with \GB. We found that in all of the simulated parameter regions where the unitarity of the model may come into question or the decay of the mediator becomes unphysical, the model is excluded by experiments that are less sensitive to the high energy behaviour of the theory. Since the model exclusion most strongly comes from direct detection experiments and relic abundance limits, the surviving parameter space is split in two by the DM annihilation channels. The overall result is a series of limits that are highly similar to, but slightly weaker than, those found for corresponding scalar and fermionic DM models in the previous study in this series~\cite{DMSimpI}.
 
\begin{figure}[tbp]
  \centering
  \includegraphics[width=\columnwidth]{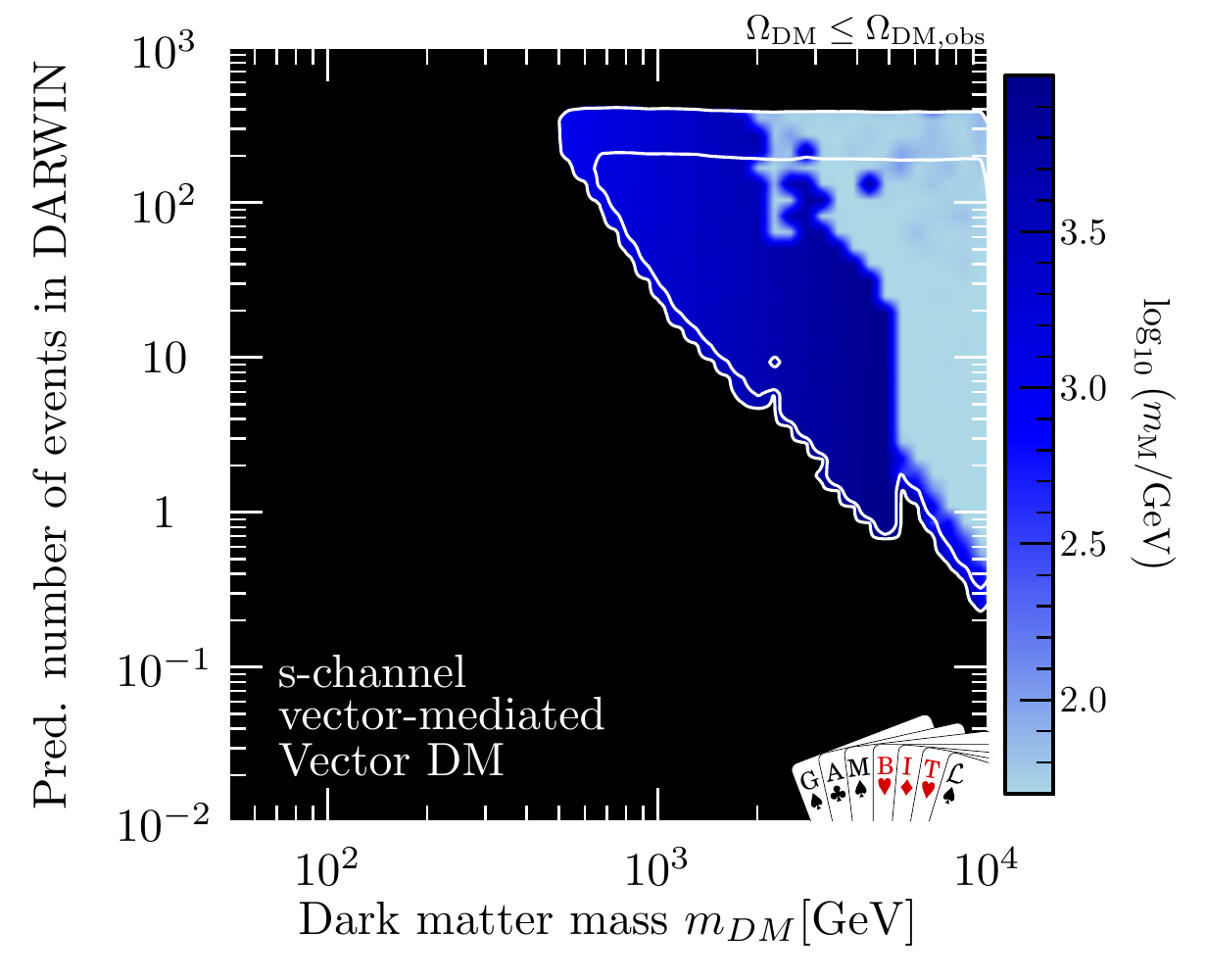}
  \caption{Predicted number of signal events in the DARWIN experiment, coloured by the mediator mass.  1$\sigma$ and 2$\sigma$ profile likelihood contours are shown in white.}
  \label{DARWIN}
\end{figure}

In the coming years, many experiments are expected to take data that may be used to constrain the model that we consider. In Figure~\ref{DARWIN} we show the predicted number of signal counts at the next-generation liquid Xenon direct detection experiment, DARWIN~\cite{DARWIN}. Within the surviving parameter space of the model, up to several hundred recoil events may be observed. Depending on how effectively the background can be rejected, a large portion of the surviving parameter space in these scans may be ruled out in the absence of any signal measurements.

We also checked the extent to which future observations by the Cherenkov Telescope Array (CTA) would constrain the model, using the same methods as in Ref.\ \cite{DMSimpI}.  None of the currently viable parameter space will be probed by CTA, with the parameter space along the resonance region far out of reach because the annihilations occur through the $p$-wave suppressed channel to quarks.

Finally, we note that further constraints can be expected from Run 3 of the LHC and the subsequent high-luminosity phase, as well as future colliders. In order to correctly interpret these constraints, it will become increasingly important to understand the high-energy behaviour of simplified models.

\section*{Acknowledgements}

CC would like to acknowledge KIT for its support and hospitality as a hosting university. This work was in part performed using the Cambridge Service for Data Driven Discovery (CSD3), part of which is operated by the University of Cambridge Research Computing on behalf of the STFC DiRAC HPC Facility (\url{www.dirac.ac.uk}). The DiRAC component of CSD3 was funded by BEIS capital funding via STFC capital grants ST/P002307/1 and ST/R002452/1 and STFC operations grant ST/R00689X/1. DiRAC is part of the National e-Infrastructure. PS acknowledges funding support from the Australian Research Council under Future Fellowship FT190100814. TEG and FK were funded by the Deutsche Forschungsgemeinschaft (DFG) through the Emmy Noether Grant No. KA 4662/1-1 and grant 396021762 - TRR 257. MJW is supported by the ARC Centre of Excellence for Dark Matter Particle Physics (CE200100008). This article made use of \textsf{pippi v2.2} \cite{pippi}.

\appendix
\setcounter{table}{0}
\renewcommand\thetable{A\arabic{table}}
\renewcommand\thetocsection{\Alph{section}}

\onecolumn
\section{Unitarity Bound including $b_{6}$ and $b_7$ couplings}
\label{section:appendix}

If the $b_6$ and $b_7$ couplings from eq.~\eqref{ModelEqFull} are allowed to be nonzero, the unitarity bound becomes
\begin{align}
\begin{split}
     \frac{1 }{32 \pi m_{\text{DM}}^4 (\frac{s}{2} - 2 m_{\text{DM}}^2)^3}  \sqrt{\frac{s - 4 m_{\text{DM}}^2}{s}}   \Bigg| \frac{b_{5}^2 }{3} \Bigg[ 2 \Big((\frac{s}{2} - 2 m_{\text{DM}}^2)^2 \big(\frac{s^2}{4} ( m_{\text{DM}}^2 - s ) + \frac{3 m_{\text{DM}}^2 s}{4}(3s - 4 m_{\text{DM}}^2) \\
     + 3m_{\text{DM}}^4 (\frac{s}{2} - 2m_{\text{DM}}^2) \big) + \frac{3 m_{\text{M}}^2 s}{8} (s -4  m_{\text{DM}}^2)  (\frac{3s^2}{8} + \frac{1}{2} m_{\text{DM}}^2 s - 4m_{\text{DM}}^4) + \frac{3m_{\text{M}}^4 s^2}{32} (s - 4 m_{\text{DM}}^2) \Big)\\
    - 3 \big(\frac{m_{\text{M}}^2 s}{4}+m_{\text{DM}}^2 (\frac{s}{2} - 2 m_{\text{DM}}^2) \big)^2 \big(2s - 4 m_{\text{DM}}^2 + m_{\text{M}}^2  \big) \ln{\Big(\frac{s - 4 m_{\text{DM}}^2+m_{\text{M}}^2}{m_{\text{M}}^2} \Big)}  \Bigg] \\
        + \frac{ b_{5} \text{Im}(b_{6})  s }{12 } \Bigg[  - (2 s - 8 m_{\text{DM}}^2) \Big( (\frac{s}{2} - 2 m_{\text{DM}}^2)^2 (  -  2 s + 9m_{\text{DM}}^2) - 3 m_{\text{M}}^2 (\frac{s}{2} - 2 m_{\text{DM}}^2)  ( -\frac{3s}{4} - m_{\text{DM}}^2) +  \frac{3 m_{\text{M}}^4 s}{4}  \Big) \\
        + 6 m_{\text{M}}^2 \big( 2 s - 8 m_{\text{DM}}^2 + m_{\text{M}}^2\big)\big( m_{\text{DM}}^2 (\frac{s}{2} - 2 m_{\text{DM}}^2) + \frac{m_{\text{M}}^2 s}{4} \big)\ln{\Big(\frac{s - 4 m_{\text{DM}}^2 + m_{\text{M}}^2}{m_{\text{M}}^2}\Big)}  \Bigg]  \\
        - \frac{\text{Re}(b_{6})^2   }{24 } \Bigg[   s (2 s - 8 m_{\text{DM}}^2) \Big( s(\frac{s}{2} - 2 m_{\text{DM}}^2)^2 - 3  \frac{m_{\text{M}}^2 s}{4} (\frac{s}{2} - 2 m_{\text{DM}}^2)  +  \frac{3 m_{\text{M}}^4 s}{4} \Big)    +  \frac{9 m_{\text{M}}^6 s^2}{2}  \ln{\Big( \frac{s - 4 m_{\text{DM}}^2 + m_{\text{M}}^2}{m_{\text{M}}^2}\Big)} \\
        +  \frac{6 s^2}{m_{\text{M}}^2} \big(\frac{s}{2} - 2 m_{\text{DM}}^2 \big)^4  - 4 s^2 \big(\frac{s}{2} - 2 m_{\text{DM}}^2 \big)^3 + 12  m_{\text{M}}^2 s^2 \big(\frac{s}{4} - m_{\text{DM}}^2 \big)^2  - 6  m_{\text{M}}^4 s^2 \big(\frac{s}{4} - m_{\text{DM}}^2 \big) \Bigg] \\
        - \frac{\text{Im}(b_{6})^2  s }{24 } \Bigg[  (2 s - 8 m_{\text{DM}}^2) \Big((\frac{s}{2} - 2 m_{\text{DM}}^2)^2(2 s -  6m_{\text{DM}}^2) - 3 m_{\text{M}}^2 (\frac{s}{2} - 2 m_{\text{DM}}^2)  (  \frac{3s}{4} -  2m_{\text{DM}}^2 ) -  \frac{3 m_{\text{M}}^4 s}{4} \Big) \\
        + 6 m_{\text{M}}^4 \big(  (\frac{s}{2} - 2 m_{\text{DM}}^2) (s -  2m_{\text{DM}}^2) + \frac{m_{\text{M}}^2 s}{4} \big) \ln{\Big(\frac{s - 4 m_{\text{DM}}^2 + m_{\text{M}}^2}{m_{\text{M}}^2}\Big)} \Bigg] \\
        - \text{Re}(b_{7})^2 m_{\text{DM}}^4 s   \Bigg[  2  (\frac{s}{2} - 2 m_{\text{DM}}^2) \big( \frac{s}{2} - 2 m_{\text{DM}}^2   +  m_{\text{M}}^2  \big)  + m_{\text{M}}^2 \big( -s  + 4 m_{\text{DM}}^2 - m_{\text{M}}^2   \big) \ln{\Big( \frac{s - 4 m_{\text{DM}}^2 + m_{\text{M}}^2}{m_{\text{M}}^2}\Big)} \Bigg]  \Bigg| \leq \frac{1}{2} \,,
\end{split}
\end{align}
where $b_{5}$ corresponds to the coupling $g_{\text{DM}}$ in the model we adopt. For the proof of the relation, we refer the reader to the supplementary Zenodo record for this study~\cite{Zenodo_DMsimpII}. The $b_{6}$ and $b_{7}$ couplings are split into their real and imaginary components, with the CP-violating couplings left in for completion. The imaginary component of the $b_{7}$ cancels in the formation of the bound. In the limit of high $s$, this simplifies to
\begin{align}
        \Bigg| - \frac{b_{5}^2 s^2 }{96 \pi m_{\text{DM}}^4 }  + \frac{ b_{5} \text{Im}(b_{6})  s^2 }{48 \pi m_{\text{DM}}^4 } - \frac{\text{Re}(b_{6})^2 s^3}{256 \pi m_{\text{M}}^2 m_{\text{DM}}^4 }   - \frac{\text{Im}(b_{6})^2  s^2 }{96 \pi m_{\text{DM}}^4}   - \frac{ \text{Re}(b_{7})^2  }{8 \pi  } \Bigg| \leq \frac{1}{2} \,.
\end{align}
The term from the real component of the $b_{7}$ coupling is independent of $s$.
\twocolumn

\bibliography{R3}

\end{document}